\documentclass[11pt,a4paper]{article}
\usepackage{amsmath,amssymb,ascmac,amsthm,bm} 
\usepackage[all]{xy}
\usepackage[dvips]{graphicx} 
\usepackage{mathrsfs} 

\usepackage{makeidx}
\makeindex

\def\slash#1{\setbox0=\hbox{$#1$}                 
   \dimen0=\wd0                                 
   \setbox1=\hbox{/} \dimen1=\wd1               
   \ifdim\dimen0>\dimen1                        
      \rlap{\hbox to \dimen0{\hfil/\hfil}}      
      #1                                        
   \else                                        
      \rlap{\hbox to \dimen1{\hfil$#1$\hfil}}   
      /                                         
   \fi}                                         %
   
\def\rc#1{\textcolor{red}{#1}}
\def\erase#1{}

\begin{document}
\null \hfill Preprint TU-908  \\[3em]
\begin{center}
{\large\bf D-branes in Generalized Geometry and Dirac-Born-Infeld Action}\\[2em] 
\end{center}

\begin{center}
{T. Asakawa\footnote{
e-mail: asakawa@tuhep.phys.tohoku.ac.jp}, S. Sasa\footnote{
e-mail: sasa@tuhep.phys.tohoku.ac.jp}, and S. Watamura\footnote{
e-mail: watamura@tuhep.phys.tohoku.ac.jp}}\\[3em] 

Particle Theory and Cosmology Group \\
Department of Physics \\
Graduate School of Science \\
Tohoku University \\
Aoba-ku, Sendai 980-8578, Japan \\ [5ex]
\end{center}

\vskip 2cm

\numberwithin{equation}{section}

\abstract
The purpose of this paper is to formulate the Dirac-Born-Infeld (DBI) action 
in a framework of generalized geometry and clarify its symmetry. 
A D-brane is defined as a Dirac structure 
where scalar fields and gauge field are treated on an equal footing in a static gauge. 
We derive generalized Lie derivatives corresponding 
to the diffeomorphism and B-field gauge transformations 
and show that the DBI action is invariant under non-linearly 
realized symmetries for all types of diffeomorphisms 
and B-field gauge transformations. 
Consequently, we can interpret not only the scalar field but also the gauge field on the
D-brane as the generalized Nambu-Goldstone boson.


\eject

\tableofcontents


\section{Introduction}
\label{sec:intro}

It is known that the low-energy effective theory for a single D-brane in slowly varying approximation is 
described by the Dirac-Born-Infeld (DBI) action.
Although it is derived from 
 the analysis of disk amplitudes in string worldsheet theory \cite{Fradkin:1985qd,Abouelsaood:1986gd,Callan:1988wz,Tseytlin:1997csa,Tseytlin:1999dj}, it is not obvious why such an action appears from the target space viewpoint.
It is an interesting question whether the DBI action can be characterized from the geometrical set up 
and symmetry principle without referring to string theory.

In the field theory, the spontaneous symmetry breaking and its non-linear realization 
is a powerful method to determine a low energy effective action.
In the presence of an extended object like a D-brane, 
 the invariance under the Poincar\'e transformation in the target Minkowski space is broken, and the scalar fields describing transverse displacements 
can be identified as Nambu-Goldstone (NG) bosons \cite{Nambu:1960xd,Goldstone:1961eq} for the broken translational symmetries.
The full Poincar\'e group symmetry is then non-linearly realized on the scalar fields, 
and their effective theory is governed by the Nambu-Goto action with derivative corrections \cite{Low:2001bw,Aharony:2011gb,Dubovsky:2012sh,Gomis:2012ki}.

Recently, this kind of argument is extended to include a $U(1)$ gauge field on a D-brane.
In \cite{Gliozzi:2011hj}, the transformation law of the gauge fields 
under the full Poincar\'e symmetry was found, 
and it was argued that the DBI action is the unique invariant 
term under the broken Lorentz symmetry in the lowest approximation. 
In \cite{Casalbuoni:2011fq}, this transformation law 
was explained using compensating diffeomorphisms to keep the static gauge, 
and it was shown that the gauge field is a covariant field under the broken Poincar\'e symmetry.
Thus, in their approach the gauge field does not appear as a NG boson and
in this sense it does not explain why gauge fields should appear in the low energy theory.

In this paper, we formulate the D-brane in the framework of
the generalized geometry, we show that not only the scalar field but also
 the gauge field on the D-brane naturally appears as a NG boson, 
and the DBI action is characterized as a generalization of the Nambu-Goto action.

The basic idea is quite simple. 
Since a T-duality transformation exchanges 
a scalar field and a component of a gauge field,
we would expect that the latter should also be a NG boson for some broken symmetry.
On the other hand, T-duality mixes the metric and the B-field in the bulk, 
as well as their associated 
symmetries (generalized isometry).
Thus, these considerations suggest 
that a gauge field is a NG boson for a spontaneously broken 
gauge transformation for a B-field.
To formulate an effective theory following this idea, 
we need an appropriate framework to incorporate the properties 
of T-duality together with 
the symmetry and its spontaneous breaking into a geometrical picture.
The generalized geometry proposed by Hitchin provides such a framework \cite{Hitchin:2004ut}.

The generalized geometry is a generalization of differential geometry,
in which a pair consisting of a tangent and a cotangent bundle is regarded as a single generalized tangent bundle.
As a result, vector fields and 1-forms are combined into generalized vector fields,
where the Lie bracket of vector fields is generalized to the Courant bracket.
Despite of the simplicity of the idea, 
this generalization unifies various distinct structures as follows: a generalized complex structure unifies both a complex structure and a symplectic structure, and a generalized Riemannian structure unifies a Riemannian metric and a B-field. 
These unifications are the reflection of the properties of the closed string
and the T-duality in superstring theory.

Several approaches have been proposed that describe D-branes in the framework 
of the generalized geometry \cite{Gualtieri:2003dx,Zabzine:2004dp,Grange:2005nm,Hatsuda:2012uk},
(see also \cite{Koerber:2010bx}, a nice review on this subject.).
However, 
they are not sufficient for our purpose.
Thus, we need to develop a formulation of D-branes in generalized geometry further,
and that is another purpose of this paper.
Here, we seek for a formulation where we do not need to impose 
any extra condition from string theory,
like T-duality, by hand.
In such a formulation, all these stringy informations should be built 
in the geometrical framework a priori. The advantage of such an approach
shed some light on the properties of the effective theory as we see in the following.

This paper is organized as follows. 
 The first part of this paper 
is devoted to the geometrical formulation of 
the D-brane valid for arbitrary spacetimes, not restricted to Minkowski spacetime.
After a preparation for some facts on the generalized geometry in 
section 2, we start with introducing a Dirac structure 
as a local description of a D-brane in section 3, 
where the gauge field and the scalar fields are treated on an equal footing.
We emphasize that this structure characterizes a spacetime 
admitting D-branes independent of closed string structure as a generalized Riemannian structure.
Then we study the symmetry transformation 
of ${\rm Diff}(M)\ltimes \Omega^2_{\rm closed}(M)$ 
associated with the generalized tangent bundle 
and its action on the Dirac structure in section 4.
Using this formulation, we derive the non-linear transformation law 
for gauge and scalar fields in a purely geometric way.
This non-linear transformation law will lead us to the 
interpretation of the gauge field as a Nambu-Goldstone boson 
for broken B-field gauge transformations.
Restricting to the 
Poincar\'e symmetry, our result coincides 
with the non-linear transformation found in \cite{Gliozzi:2011hj}.

Next, in addition to the Dirac structure, 
a generalized Riemannian structure is considered in section 5.
By introducing the notion of the metric seen by the Dirac structure,
remarkably, we obtain a Buscher-like rule of a generalized metric without using 
T-duality.

In section 6, 
we study the invariance of the DBI action under the non-linear 
transformation given in this paper. We also discuss how the DBI action
is characterized by the full symmetry. By specializing to the Minkowski spacetime, 
this analysis also shows the difference between broken translational symmetries 
and broken Lorentz symmetries, where the NG bosons 
appear associated only with the former. We also conclude that the
gauge field is interpreted as a Nambu-Goldstone boson.

\section{Preliminary}
\label{sec:preliminary}
\hspace{5.2mm} In this section, we briefly review basic facts on 
generalized geometry, proposed originally by Hitchin \cite{Hitchin:2004ut} and developed by Gualtieri \cite{Gualtieri:2003dx}
and introduce notations used in this paper.
Further details are found in \cite{Gualtieri:2007ng,Hitchin:2010qz,roytenberg-1999}. 
Review articles for physicists are for example in \cite{Zabzine:2006uz,Bouwknegt:2010zz,Koerber:2010bx}.

\subsection{Generalized tangent bundle}
\label{subsec:gene_tan_bndl:g4}
\hspace{5.2mm} Let $M$ be a smooth $D$-dimensional manifold 
corresponding to a target spacetime. 
A generalized tangent bundle
over $M$,
\begin{equation}
{\mathbb T} M = TM \oplus T^\ast M
\end{equation}
is a sum of the corresponding tangent bundle and cotangent bundle.
We denote 
a section of a generalized bundle as a formal sum $v+\xi \in \Gamma({\mathbb T}M)$, 
where $v=v^M\partial_M \in \Gamma(TM)$ $(M=0, \cdots, D-1)$ is a vector field and 
$\xi=\xi_M dx^M \in \Gamma(T^\ast M)$ is a differential 1-form. 
The space of sections, $\Gamma({\mathbb T} M)$ is equipped with
\begin{itemize}
\item[-] an anchor map $\pi: \Gamma({\mathbb T} M)\to \Gamma(TM)$, given by a projection onto vector fields 
\begin{align}
\pi(v+\xi)=v,
\end{align}
\item[-] a fiberwise non-degenerate symmetric bilinear form (canonical inner product)
\begin{equation}
\left< u+\xi , v + \eta \right> = \frac{1}{2} (\imath_u \eta + \imath_v \xi) 
=\frac{1}{2} \begin{pmatrix} u \\ \xi \end{pmatrix}^T \begin{pmatrix}0&1\\ 1&0 \end{pmatrix} \begin{pmatrix}v \\ \eta \end{pmatrix},
\label{eq:canonical_inner_product:3kdo}
\end{equation}
\item[-] the Dorfman bracket
\begin{equation}
[u+\xi , v+\eta] = [u, v] + {\mathcal L}_u \eta - \imath_v d\xi,
\label{eq:Dorfman_bracket:die4t}
\end{equation}
where the first term in the r.h.s. is the ordinary Lie bracket of vector fields, 
$\imath_u$ is an interior product , i.e. $\imath_u \eta = u^M \eta_M$ and ${\mathcal L}_u $ is the Lie derivative along a vector field $u$.
\end{itemize}

These structures (together with their compatibility conditions) make $\Gamma({\mathbb T} M)$ a Courant algebroid \cite{MR998124,Liu:1997fj}.
It is a natural generalization of the Lie algebroid structure on vector fields $\Gamma(TM)$. 
We sometimes abbreviate the symbol $\Gamma({\mathbb T} M)$ as ${\mathbb T} M$ in the following.

Due to the canonical inner product (\ref{eq:canonical_inner_product:3kdo}), the generalized tangent bundle has the structure group $O(D,D)$. A structure group $GL(D)$ of $TM$ is a subgroup of $O(D,D)$.

Note that we work with the Dorfman bracket rather than the Courant bracket 
which is the anti-symmetrization of the Dorfman bracket.
In the presence of closed $3$-form ($H$-flux), it is known that the Courant algebroid structure 
is modified either by replacing the bracket by its H-twisted version, or by twisting the  
generalized tangent bundle glued by using B-field transformation (see below) in addition to diffeomorphism. 
In this paper, we will concentrate on the case of vanishing $H$-flux for simplicity, 
but it is possible to include an $H$-flux.

\subsection{Symmetry of the Courant algebroid}
\label{subsec:Gene_Lie_der:9f4}
\hspace{5.2mm} As the tangent bundle $TM$ has the diffeomorphism ${\rm Diff}(M)$ as its symmetry, the generalized tangent bundle possesses 
the symmetry ${\rm Diff}(M)\ltimes \Omega^2_{\rm closed}(M)$, 
a semi-direct product of the 
following two transformations:
\begin{itemize}
\item[-] ${\rm Diff}(M)$: For a diffeomorphism $f: M\to M$ of the base manifold, 
\begin{equation}
f_\ast \oplus f^{\ast -1} : {\mathbb T} M \to {\mathbb T} M , \qquad u+\xi \mapsto f_\ast (u) + f^{\ast -1} (\xi),
\end{equation}
is induced on ${\mathbb T} M$, where $f_\ast : TM \to TM$ and $f^\ast : T^\ast M \to T^\ast M$ are a pushforward and a pullback, respectively. It is called a generalized pushforward and  denoted by the same symbol $f_\ast = f_\ast \oplus f^{\ast -1}$.
\item[-] B-field transformation: 
For a closed 2-form $B \in \Omega^2_{\rm closed}(M)$, it is defined as
\begin{equation}
e^B : {\mathbb T} M \to {\mathbb T} M , \qquad u+\xi \mapsto u+\xi+ \imath_u B,
\end{equation}
which shift a 1-form. 
\end{itemize}
It is shown that these two transformations form 
the automorphism group of the Courant algebroid \cite{Gualtieri:2003dx}.
The corresponding infinitesimal version, the derivation ${\rm Der}({\mathbb T} M)$, 
is generated by a pair of a vector field and a $2$-form 
$(X,B) \in TM\oplus \wedge^2 T^* M$. 
A generalized Lie derivative ${\cal L}_{(X,B)}$ acting on ${\mathbb T}M$ is defined as \cite{Hu}
\begin{equation}
{\cal L}_{(X,B)} (u+\xi) ={\cal L}_X (u + \xi ) + \imath_u B,
\end{equation}
where ${\cal L}_X$ is the ordinary Lie derivative.
Note that if $B=-d\Lambda$ is an exact $2$-form written by a $1$-form $\Lambda$, 
the above generalized Lie derivative reduces to the Dorfman bracket. 
We denote this case by ${\cal L}_{(X,-d\Lambda)} (u+\xi) ={\cal L}_{X+\Lambda} (u+\xi)=
[X+\Lambda, u+\xi]$.
Thus, the Dorfman bracket (\ref{eq:Dorfman_bracket:die4t}) is a sum of an ordinary Lie derivative (first two terms) and a NS-NS B-field gauge transformation (last term).\\

\subsection{Dirac structure}
\label{subsec:Dircstr:3dl}

\hspace{4.2mm} 
A skew-symmetry is failure in the Dorfman bracket  due to the second and third terms of (\ref{eq:Dorfman_bracket:die4t}). We can find a subbundle in which the Dorfman bracket becomes the Lie bracket.

A Dirac structure \cite{MR998124} is defined as a subbundle $L \subset {\mathbb T}M$ of rank $D$ such that
\begin{itemize}
\item[-] isotropic: $L$ is self-orthogonal $L=L^\perp$, i.e. $\left< a,b \right>= 0$ for $^\forall a, b \in \Gamma(L)$
\item[-] involutive: $L$ is closed under the Dorfman bracket, i.e. $[a,b]\in \Gamma(L)$ for $^\forall a, b \in \Gamma(L)$.
\end{itemize}
Obviously, a Dirac subbundle $L$ has the structure of a Lie algebroid, 
where the anchor map $\rho: L\to TM$ is $\rho=\pi \circ \iota$ 
with $\iota$ being the inclusion of $L$ into ${\mathbb T}M$.
For a general theory of Lie algebroid, see \cite{MR896907}.
The simplest examples of Dirac structures are $TM$ with $\rho={\rm id.}$ 
and $T^\ast M$ with $\rho=0$.
It is known that any Lie algebroid defines a (singular) foliation. 

Note that if both a subbundle $L \subset {\mathbb T}M$ and 
its dual bundle $L^\ast$ are Dirac structures 
and the generalized tangent bundle ${\mathbb T} M$ 
splits as 
${\mathbb T} M = L \oplus L^\ast$, it is called a generalized product structure \cite{Zabzine:2004dp,Koerber:2010bx}. 
Historically, the notion of Courant algebroid, given in \cite{Liu:1997fj} with its primary example, is a Lie bialgebroid $A\oplus A^*$ given by a pair of Lie algebroids $A$ and $A^*$.

A more familiar structure is a generalized complex structure, where a subbundle and its dual belong to the complexified generalized tangent bundle ${\mathbb T}^{\mathbb C} M$, but we do not use it in this article.

\subsection{Generalized Riemannian structure}
\label{sec:gene_Riem}
\hspace{5.2mm} A generalized Riemannian structure (generalized metric) is defined as a positive definite subbundle $C_+ \subset {\mathbb T} M$ of rank $D$ of the generalized tangent bundle, i.e., the canonical inner product restricted to $C_+$ is positive definite,
$\left< A, A \right> > 0$ for all non-zero sections $A \in \Gamma(C_+)$.
By defining $C_-$ as the orthogonal complement of $C_+$, which is a negative definite subbundle, 
the generalized tangent bundle splits as ${\mathbb T} M= C_+ \oplus C_-$. 
Thus specifying the generalized metric $C_+$ is equivalent to a reduction of the structure group from $O(D,D)$ to $O(D)\times O(D)$.

Since $C_+ \cap T^\ast M = \{ 0 \}$, the generalized metric can be also written as the graph of the map $E=g+B : TM \to T^\ast M$, namely $C_+ $ is written as
\begin{equation}
C_+=\{ V_+=  v + (g+B)(v)\,|\,v\in TM\},
\label{eq:Genemetricasgraph:3kdo}
\end{equation}
where $g$ is an ordinary Riemannian metric and $B$ is a two-form identified as a NS-NS B-field. 
Indeed, one has $\left< V_+, V'_+ \right>=g(v,v')$ and $C_+$ is positive definite,
i.e., the canonical inner product
 restricted to $C_+$ defines an ordinary Riemannian structure.
Correspondingly, the negative-definite subbundle $C_-$ is given by the graph of the map $-g+B$ as
\begin{equation}
C_- =\{ V_-=  v + (-g+B)(v) \,|\,v\in TM\}.
\label{eq:negaGenemetricasgraph:3k0f}
\end{equation}

The generalized Riemannian structure is also specified by a self-adjoint orthogonal endomorphism $G: {\mathbb T} M \to {\mathbb T} M$ such that 
$G^2=G^TG=1$ and 
$\left< A, GA \right> > 0$
 for all $A \not= 0 \in \Gamma({\mathbb T} M)$. 
Then, the subbundles $C_{\pm}$ are expressed as the $\pm1$-eigenspaces of $G$, i.e. $C_{\pm} = {\rm Ker}(1\mp G)$. 
Solving the eigenvalue equation
\begin{equation}
GV_{\pm}= \pm V_{\pm} \label{eq:eigenvalueeq:8hst}
\end{equation}
for $V_{\pm} \in \Gamma (C_\pm)$ given in  (\ref{eq:Genemetricasgraph:3kdo}), (\ref{eq:negaGenemetricasgraph:3k0f}), the generalized metric $G$ is given by
\begin{equation}
G = \begin{pmatrix} -g^{-1}B & g^{-1} \\ g-Bg^{-1}B & Bg^{-1} \end{pmatrix}
\label{eq:indgenemetG_oscj9d}
\end{equation}
as a matrix of a map $\Gamma(TM \oplus T^\ast M) \to \Gamma(TM \oplus T^\ast M)$.

As we have seen, there are various ways to characterize a generalized Riemannian structure, 
that are mathematically equivalent. However, physically it is not apparent 
which part of the generalized metric such as $g$, $g-Bg^{-1}B$, or $G$ should be used in a Lagrangian or a Hamiltonian of the effective theory.
In a first quantized string, or in the double field theory \cite{Hull:2009mi}, $G$ plays a primary role.
In this paper, after studying D-branes without imposing this structure, 
we give another way to describe a generalized metric suitable for D-branes.

\section{D-branes as Dirac structures}
\label{sec:D-brane_Dirac:e4g}
\hspace{5.2mm}
In this section, we describe a D-brane as a Dirac structure, 
in which the scalar fields and a gauge field on the D-brane are treated on an equal footing.

Starting with a conventional description of a D-brane as an embedding of a worldvolume into a target spacetime, we rewrite it as a leaf of a foliation of that spacetime.
This description is easily generalized to incorporate a gauge field 
by using a diffeomorphism and a B-field gauge transformation.
We will see that this process can be also understood as a construction using a generalized connection.

\subsection{Embedding of a D-brane and foliations}
\label{subsec:foliation}
\hspace{5.2mm} Let us first describe the conventional picture for a D-brane.
A worldvolume $\Sigma$ of a D$p$-brane is a ($p+1$)-dimensional manifold 
embedded into a $D$-dimensional target space $M$ by a map $\varphi : 
\Sigma \hookrightarrow M$. In addition, a D-brane is associated 
with a complex line bundle $V\to \Sigma$ 
with a connection. 
The possible embedding maps and the possible choices of connections are 
regarded as the dynamical degrees of freedom associated to a D-brane.
Here we focus on embedding maps and reformulate them 
in a form such that the relation to the generalized geometry is apparent.

Here, we take the worldvolume $\Sigma={\mathbb R}^{p+1}$ and a target 
space $M={\mathbb R}^D$ as Euclidean spaces for simplicity,
but our argument is applicable to any curved manifold 
if we consider its local structure.  

Let us recall the notion of the static gauge in some detail.
For given coordinates $\sigma^a$ ($a= 0, \cdots, p$) on a worldvolume $\Sigma$
and $x^M$ ($M= 0, \cdots D-1$) on a target space $M$, a map $\varphi : \Sigma \hookrightarrow M$ is specified by $D$ functions $x^M(\sigma)$ on $\Sigma$.
Among them, we fix a reference map $\varphi$ as $(x^a(\sigma), x^i (\sigma))=(\sigma^a, 0)$ ($a= 0, \cdots, p$, $i= p+1, \cdots, D-1$) such that the submanifold $\varphi(\Sigma)$ is a hypersurface in $M$ represented by $x^i=0$, which is called the static gauge%
\footnote{
Note that it is always possible locally for any manifold $M$:
For a point on a submanifold $p \in \varphi(\Sigma)$, one can choose a open cover $\{U_\alpha\}$ of $M$ and local coordinates $x^M$ such that $\varphi(\Sigma)\cap U_\alpha$ is represented by $(x^a, x^i=0)$ in the neighborhood of $p$. 
Then, by using a diffeomorphism on the worldvolume $\Sigma$, we can choose coordinates $\sigma^a$ on $\Sigma$ as $\sigma^a=x^a$.}.
This enables us to identify the coordinates of the worldvolume $\Sigma$ and the embedded submanifold $\varphi(\Sigma)$, i.e., $x^a=\sigma^a$.
Note that the map $\varphi$ induces a pushforward $(d\varphi)_p :T_p \Sigma \to T_{\varphi(p)}M$ on corresponding tangent spaces.
A vector $v^a\partial/\partial \sigma^a \in T_p \Sigma$ at a point $p\in \Sigma$ also maps to a vector $v^a \partial_a \in T_{\varphi(p)}M$ at $\varphi \in M$ with abbreviation $\partial_a =\partial/\partial x^a$.

A D-brane can fluctuate around a fixed configuration.
In the static gauge, such dynamical degrees of freedom are transverse displacements represented by scalar fields $\Phi^i(\sigma)$ on the worldvolume $\Sigma$. 
They give a new embedding map
\begin{equation}
\varphi_\Phi :\Sigma \hookrightarrow M , \ \ \ \ \  (x^a(\sigma), x^i (\sigma))=(\sigma^a, \Phi^i(\sigma)).
\end{equation}
It is equivalent to a new hypersurface $x^i=\Phi^i(x^a)$ in $M$.
Similarly, its pushforward is  
\begin{equation}
(d\varphi_\Phi)_p : T_p\Sigma \to T_{\varphi_\Phi (p)}M, \ \ \ \ \  
v^a \frac{\partial}{\partial \sigma^a} 
\mapsto v^a (\partial_a+\partial_a\Phi^i \partial_i).
\end{equation}
The scalar fields determine both the 
position of the D-brane as well as the basis of tangent vectors.

\begin{figure}[tbp]
\begin{center}
\begin{minipage}{10cm}
\includegraphics[scale=0.6,width=10cm]{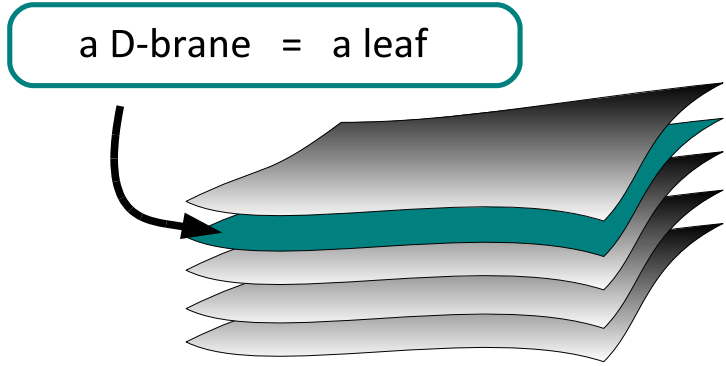}
\caption{\footnotesize Schematic picture of the foliated structure, where 
a D-brane is identified with a certain leaf.
}
\label{fig:Dirac_str}
\end{minipage}
\end{center}
\end{figure}

We would like to rewrite these settings purely on the target space $M$, without recourse to the worldvolume $\Sigma$.
For this, it is useful to think of a foliation of the target space $M$, by considering infinitely many copies of submanifolds simultaneously (see Fig.\ref{fig:Dirac_str}).
A submanifold $\varphi(\Sigma)\simeq \Sigma$ wrapped by a D-brane is considered as a leaf of this foliation.
It is equivalent to specify a subbundle $\Delta \subset TM$ over $M$, where $\Delta =\{ v^a (x) \partial_a\}$ is a set of vector field tangent to leaf directions%
\footnote{$(d\varphi)_p$ is extended to a pushforward of vector fields 
$\varphi_* :T \Sigma \to TM|_{\varphi(\Sigma)}$, when restricted on the submanifold, but it is not defined on the whole $M$.}.
Its restriction to a particular leaf is $\Delta|_{\Sigma}=T\Sigma$.
For this we demand that the global existence of the foliation structure, 
which is equivalent to the condition that the submanifold $\Sigma$ is a leaf of the foliation and is an integrable submanifold of $M$. 
In the case of the Minkowski spacetime, this assumption is valid for any point $p$.
Similarly, the map $\varphi_\Phi$ defines another foliation on $M$.

Now the change of embedding caused by the scalar fields is 
written in the target space $M$, without recourse to the worldvolume $\Sigma$.
Let $\Phi = \Phi^i (x^a) \partial_i \in TM$ be a vector field defined on the whole target space $M$, whose coefficient functions are $x^i$-independent%
\footnote{We sometimes call a vector field $\Phi=\Phi^i\partial_i$ made out of scalar fields $\Phi^i$ as ``scalar fields". 
This terminology originates form the viewpoint of the worldvolume $\Sigma$. Do not confuse !}.
It generates a diffeomorphism as a Lie derivative $-{\cal L}_\Phi$.
Its 
action on the subbundle $\Delta$ is 
\begin{equation}
e^{-{\cal L}_\Phi}v^a(x^a, x^i)\partial_a = 
v^a(x^a, x^i-\Phi^i(x^a))(\partial_a+ \partial_a\Phi^i \partial_i),
\end{equation}
which relates two subbundles $\Delta={\rm span}\{\partial_a\}$ and $e^{-{\cal L}_\Phi}\Delta={\rm span}\{\partial_a +\partial_a\Phi^i \partial_i\}$%
\footnote{Note the minus sign in the argument. 
If $v^a(x^a, x^i)$ is a function on $M$ peaked at $x^i=0$,
$v^a(x^a, x^i-\Phi^i(x^a))$ has a peak at $x^i=\Phi^i(x^a)$.}

\begin{figure}[tbp]
\begin{center}
\begin{minipage}{10cm}
\includegraphics[scale=0.50]{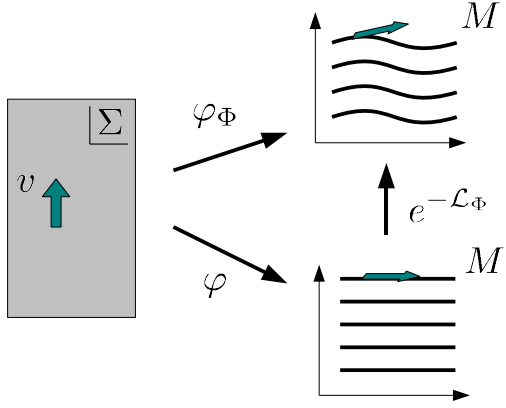}
\caption{\footnotesize Fluctuations can be described by using a diffeomorphism generated by a
vector field defined by the scalar field $\Phi^i$.}
\label{fig:diff_foliation}
\end{minipage}
\end{center}
\end{figure}

In summary, a D-brane is described as a leaf of a foliation,
and a transverse fluctuation of a D-brane corresponds to a deformation of the foliation.
This target space picture of D-branes is easily extended 
in terms of generalized geometry
to include gauge fields as we will see next.

\subsection{Generalized embedding and Dirac structure}
\label{subsec:embedding_D-brane:d3d}
Here we extend the previous formulation of D-brane to the framework of generalized geometry.
Given a target space $M$ with its local coordinates $(x^a,x^i)$ and its generalized tangent bundle ${\mathbb T}M$, let us define a  
subbundle $L= {\rm span}\{\partial_a, dx^i \} \subset {\mathbb T} M$ over $M$.
A section has the form
\begin{equation}
V_{L} = v^a (x)\partial_a + \xi_i (x) dx^i \quad \in L. 
\end{equation}
It is a generalization of a subbundle $\Delta\subset TM$ as $L=\Delta \oplus {\rm Ann}(\Delta)$, where ${\rm Ann}(\Delta) \subset T^*M$ denotes an annihilator of $\Delta$ in ${\mathbb T}M$
\footnote{$Ann(\Delta):=\{\xi\in\Gamma(T^*M)|\forall V\in \Gamma(\Delta),\langle\xi,V\rangle=0\} $.}.
The sections of the dual bundle $L^\ast={\rm span}\{ \partial_i, dx^a \}={\rm Ann}(\Delta)^* \oplus \Delta^*$ 
have the form
\begin{equation}
V_{L^\ast} = v^i (x)\partial_i + \xi_a (x) dx^a \quad \in L^\ast. 
\label{eq:L_0Diracstr:4k0d}
\end{equation}
It is easy to show that both subbundles $L$ and $L^*$ are Dirac structures and they define a generalized product structure $TM\oplus T^*M=L\oplus L^*$.

Conversely, a generalized product structure $TM\oplus T^*M=L\oplus L^*$ 
is a structure added on the target space that admit a D-brane as a leaf of a foliation \cite{Zabzine:2004dp}\cite{Koerber:2010bx}.
The subbundle $L$ plays the role of the static gauge, as we have discussed.
Our proposal in this paper is that all the geometric information realized on a D-brane is seen by this Dirac structure $L$.
To demonstrate this proposal, we will consider the fluctuations on the D-brane ($\Phi^i$, $A_a$), 
non-linearly realized symmetries and a metric seen by D-brane and show that they 
can be characterized geometrically by
 using the notion of the Dirac structure.

The fluctuations are also incorporated into our formulation as follows.
We can combine the scalar fields $\Phi = \Phi^i (x^a) \partial_i$ 
and the one-form gauge field $A=A_a(x^a) dx^a$ into a generalized vector $\Phi+A \in L^*$.
We assume here that their coefficient functions are $x^i$-independent.
It is then natural to consider a generalized Lie derivative ${\cal L}_{\Phi+A}$ for a diffeomorphism and a B-field gauge transformation.
By acting it on $L$, we have another subbundle $L_{\cal F}$
\begin{equation}
L_{\cal F} =e^{-{\cal L}_{\Phi+A}}L \subset {\mathbb T} M,
\label{L_F}
\end{equation}
with its sections having the form
\begin{align}
V &= v^a (x) (\partial_a + \partial_a\Phi^i \partial_i {+F_{ab}} dx^b) 
 + \xi_i (x)(dx^i - \partial_a \Phi^i dx^a),
\label{A section}
\end{align}
where $F_{ab}=\partial_aA_b-\partial_bA_a$. 
Here ${\cal F}$ in our notation $L_{\cal F}$ denotes a generalized field strength
 corresponding to a generalized vector $\Phi+A$. (see the next subsection.)

It is straightforward to verify that $L_{\cal F}$ is a Dirac structure, 
by using the Bianchi identity $dF=0$.
We will also show that the dual $L^*$ remains unchanged under this operation.
Their sum reproduces the total space of the generalized tangent bundle, $L_{\cal F}\oplus L^* = {\mathbb T}M$,
so that the generalized product structure is intact.
Therefore, fluctuations $\Phi+A$ can be regarded as a deformation of a generalized product structure.

\begin{figure}[tbp]
\begin{center}
\begin{minipage}{10cm}
\includegraphics[scale=0.50]{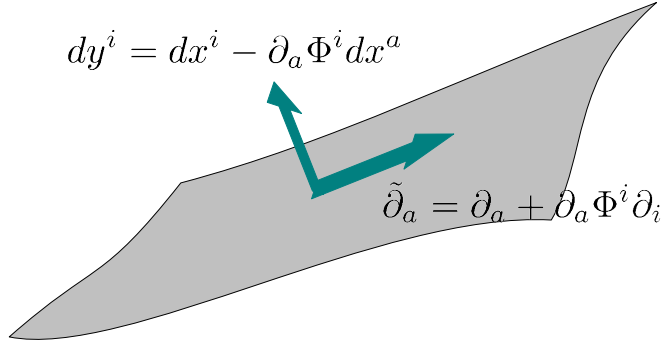}
\caption{\footnotesize The schematic picture of the basis along the leaf $L_{\cal F}$ and 
normal to the leaf.}
\label{fig:Direction}
\end{minipage}
\end{center}
\end{figure}

A Dirac structure is automatically a Lie algebroid.
For our $L_{\cal F}$, its structure is specified by the basis 
$e_a :=\partial_a+ \partial_a\Phi^i \partial_i + F_{ab} dx^b$ and 
$e^i:=dx^i - \partial_a \Phi^i dx^a$ as
\begin{align}
&\rho (e_a) =\partial_a+ \partial_a\Phi^i \partial_i, \quad \rho (e^i)=0,\\
&[e_a, e_b]=[e_a, e^i ]=[e^i, e^j ]=0,
\end{align}
where the bracket of the generalized vectors is understood to be the Dorfman bracket.
Note that this Lie bracket relation is the same as that of $L$ where the fluctuations are absent ($\Phi+A=0$).
This is a special kind of Lie algebroid with vanishing structure functions,
which means that a deformed leaf of $L_{\cal F}$ is still an integrable submanifold. 
In this case, the basis of the Lie algebroid can be represented by a coordinate basis.
Namely, if we define new coordinates $y^a=x^a$ and $y^i=x^i-\Phi^i(x^a)$, 
where the D-brane is specified by $y^i=0$, then the basis of the algebroid can be written by 
\begin{align}
\frac{\partial}{\partial y^a}:=\tilde{\partial}_a =\partial_a+ \partial_a\Phi^i \partial_i, \quad
dy^i=dx^i -d\Phi^i,
\label{tilde system}
\end{align}
At each point $p \in M$, the former span the directions along a leaf of $L_{\cal F}$,
while the latter span the directions normal to a leaf in $T^*_p M$. 
See fig \ref{fig:Direction}.

We close this section with a remark.
In a general target space $M$ and its covering $\{U_\alpha\}$, 
$\Phi+A$ may be defined only locally on each open set $U_\alpha$. 
They may be globally non-trivial, and we need to glue them in the 
overlapping region of an open covering.

\subsection{Generalized connections}
\label{subsec:gene_conn}
Here we show that the combination $A+\Phi \in L^*$ can also be regarded as a generalization of a connection $1$-form. 
Given a Lie algebroid $L$, one can formulate a differential calculus 
$(\Gamma(\wedge^\bullet L^*),d_L)$\cite{Liu:1997fj}.
In our case, the exterior differential is locally written as $d_L=dx^a \partial_a$.
For a complex line bundle $V\to M$, which is a $L$-module,  
a generalized connection ${\mathcal D}$ on a vector bundle $V$ is defined as a linear map
\begin{equation}
{\mathcal D} : \Gamma (V) \to \Gamma(V \otimes L^*),
\end{equation}
which satisfies the Leibniz rule 
${\mathcal D}(sf)={\mathcal D}(s)f+s\otimes (d_Lf)$ for ${}^\forall \!s\in \Gamma (V)$ and ${}^\forall \!f\in C^\infty(M)$.

Let ${\mathcal D}=d_L +A+\Phi$ be such a generalized connection (it is possible because $A+\Phi \in L^*$).
In particular, directional derivatives along basis $\partial_a$ and $dx^i$ in $\Gamma(L)$, we have 
\begin{eqnarray}
{\mathcal D}_{a} &\equiv& {\mathcal D}_{\partial_a} = \partial_a +A_a, \label{eq:gene_conn_Da:ogc7} \\
{\mathcal D}^{i} &\equiv& {\mathcal D}_{dx^i} = \Phi^i.
\end{eqnarray}
A generalized field strength is defined for $V,W \in \Gamma(L)$ as (note $[V,W]=0$)
\begin{equation}
{\mathcal F}(V,W)=[{\mathcal D}_V,{\mathcal D}_W]-{\mathcal D}_{[V,W]}.
\label{eq:gene_field_str:6d0w}
\end{equation}
Then we have 
\begin{align}
{\mathcal F}_{ab} &= F_{ab} = \partial_a A_b - \partial_b A_a, 
\quad {\mathcal F}_a^{\ j} = \partial_a \Phi^j, \quad 
\mathcal{F}^i_{\ b} = - \partial_b \Phi^i, \quad
{\mathcal F}^{ji} = 0. \label{eq:gene_fieldstrength_abelian:3kf9}
\end{align} 
or equivalently, 
\begin{equation}
{\mathcal F} = \frac{1}{2} F_{ab} dx^a \wedge dx^b + \partial_a \Phi^i dx^a \wedge \partial_i \quad \in \Gamma(\wedge^2 L^\ast).
\end{equation}

With this field strength, the Dirac structure $L_{\cal F}$ in (\ref{L_F}) is now written 
as a graph of the map ${\mathcal F} : L \to L^\ast$, that is, $L_{\cal F}=\{V+{\cal F}(V)\,|\, V\in L\}$.
In fact, a section for this graph is written for $V\in L$ as
\begin{equation}
\begin{split}
V + {\mathcal F}(V)
&=v^a (x) (\partial_a + \partial_a\Phi^i \partial_i + F_{ab}dx^b)+\xi_i (x) (dx^i - \partial_a \Phi^i dx^a)  \quad \in \Gamma(L_{\mathcal F}),
\end{split} 
\end{equation}
which coincides with (\ref{A section}).
Note that the dual $L^*$ is invariant under this deformation, i.e., $L^*_{\cal F}=L^*$, since ${\cal F}(L^*)=0$.
As is clear from this construction, the gauge field $A$ is actually a $U(1)$-connection on leaves defined by $L$, and there is a $U(1)$ gauge symmetry $A \to A +d_L \lambda$, which leads to the same Dirac structure $L_{\cal F}$.
By pulling-back this to a leaf, 
it is identified as a gauge field on a D-brane.
On the other hand, the scalar field $\Phi$ is a connection that lifts a curve in a leaf of $L$ to a curve in a leaf of $L_{\cal F}$. 

This is another construction of a Dirac structure, 
referred to as a deformation of Dirac structures.
In general, for an arbitrary Lie algebroid $L$ and a tensor ${\cal F}\in \Gamma(\wedge^2 L^*)$, the condition that a graph $L_{\cal F}=\{V+{\cal F}(L)\,|\, V\in L\}$ is a Dirac structure of a Courant algebroid $L\oplus L^*$ has been already obtained in \cite{Liu:1997fj} by $d_L {\cal F}+ \frac{1}{2} [{\cal F},{\cal F}]=0$, where the second term is a Schouten bracket in $\Gamma(\wedge^\bullet L^*)$.
It turns out that our assumption that $A_a$ and $\Phi^i$ are $x^i$-independent is too restrictive to define a possible Dirac structure, and can be relaxed.
For an arbitrary fluctuation $\Phi+A$, it reduces to two conditions $\partial_i (\partial_a \Phi^i) =0$ and 
${\tilde \partial}_a F_{bc} + {\tilde \partial}_b F_{ca} + {\tilde \partial}_c F_{ab} = 0$, 
proved in the appendix \ref{subsec:Dorfman_involutive}.
Our analysis in the following sections is valid for $\Phi+A$ satisfying these two conditions.
Note also that this construction is globally well-defined for any manifold $M$ 
as opposed to our previous description by Lie derivative, $L_{\cal F}=e^{-{\cal L}_{\Phi+A}}L$, where ${\Phi+A}$ is locally defined. For $L_{\cal F}$ to be globally defined by this description, 
we need to glue the local generalized vectors ${\Phi+A}$ 
on an overlapping region by a diffeomorphism and a $U(1)$ gauge transformation.
As a result, any Dirac structure in ${\mathbb T}M$ has the form of a B-field transformation of a foliation $L=\Delta_\Phi \oplus {\rm Ann} (\Delta_\Phi)$ as $L_F=e^{F}L$ \cite{Gualtieri:2003dx,MR998124}%
\footnote{Here $\Delta_\Phi={\rm span}\{\tilde{\partial}_a \}$.}.

In summary, a spacetime which admits a D-brane is characterized by a Dirac structure $L$, and the fluctuation on a D-brane is regarded as a deformation $L_{\cal F}$. We can formulate this deformation either by an action of a generalized Lie derivative or by a graph of a generalized curvature.
This shows the symmetric role of scalar (embedding) and gauge fields (connection).
In any case, it defines a possible ``shape" of a D-brane in the target space.
A choice of a leaf corresponds to the spontaneously symmetry breaking, that we will elaborate on next.

\section{Symmetry for Dirac structures}
\label{sec:gene_Liederiv:xie6}

In this section, we study the role of the symmetry of the target space for a Dirac structure $L_{\cal F}$ corresponding to a D-brane, in terms of generalized geometry.
We clarify the structure for the hierarchy of spontaneous symmetry breakings 
accompanied by the foliation preserving 
and leaf preserving diffeomorphism and its generalization.
Then we obtain the non-linear transformation law for scalar and gauge fields for broken symmetries.

\subsection{Symmetry preserved by a Dirac structure}
\label{subsec:symmetry}

The symmetry group of a generalized tangent bundle ${\mathbb T}M$ is ${\rm Diff}(M)\ltimes \Omega^2_{\rm closed}(M)$, a semi-direct product of diffeomorphisms and B-field transformations.
Since the closed $2$-form is exact and
in our case $M={\mathbb R}^D$, 
a B-field transformation reduces to a gauge transformation.
Thus, an infinitesimal transformation is labeled by the combination $\epsilon+\Lambda \in {\mathbb T}M$ of a vector field and a $1$-form, and it acts as a generalized Lie derivative $-{\cal L}_{\epsilon+\Lambda}$ on the generalized tangent bundle ${\mathbb T}M$.
Note that an exact 1-form $\Lambda =d\lambda $ does not generate the transformation of any symmetry.
In the following, it is also useful to decompose it w.r.t the generalized product structure ${\mathbb T}M=L\oplus L^*$ as
\begin{equation}
\begin{split}
\epsilon&=\epsilon_\parallel+\epsilon_\perp=\epsilon^M\partial_M=\epsilon^a\partial_a + \epsilon^i\partial_i \\
\Lambda&=\Lambda_\perp+\Lambda_\parallel=\Lambda_Mdx^M=\Lambda_i dx^i+\Lambda_a dx^a.
\end{split} \label{eq:infgene_generators:h4kd}
\end{equation}
Note that $\epsilon_\parallel +\Lambda_\perp \in L$ and $\epsilon_\perp +\Lambda_\parallel \in L^*$.

We would like to find a subgroup of ${\rm Diff}(M)\ltimes \Omega^2_{\rm closed}(M)$ 
that preserves $L$. i.e., $\epsilon+\Lambda$ that satisfies 
$L-{\cal L}_{\epsilon+\Lambda}L = L$.
Thus we need the action of a generalized Lie derivative ${\cal L}_{\epsilon+\Lambda}$ on a Dirac structure $L$.
For a section of the Dirac structure $L$,
\begin{equation}
V = v^a (x)\partial_a + \xi_i(x) dx^i \in \Gamma (L),
\end{equation}
the action of the generalized Lie derivative is obtained as 
\begin{equation}
\begin{split}
{\mathcal L}_{\epsilon+\Lambda} V 
&= (\epsilon^M\partial_Mv^a-v^b\partial_b\epsilon^a)\partial_a + (\epsilon^M\partial_M\xi_i+\xi_j\partial_i\epsilon^j -v^b \partial_{[b} \Lambda_{i]}) dx^i \\
&\quad -v^b\partial_b\epsilon^i\partial_i + (\xi_j\partial_a\epsilon^j -v^b \partial_{[b} \Lambda_{a]}) dx^a.
\end{split}
\label{LV}
\end{equation}
From this result, we observe that ${\mathcal L}_{\epsilon_\parallel +\Lambda_\perp} V$ always lies in $L$, that is a consequence that $L$ is involutive.
On the other hand, ${\mathcal L}_{\epsilon_\perp +\Lambda_\parallel} V$ produces in general a $L^*$-component. It keeps an element of $L$, iff $\partial_b \epsilon^i=0$ and $\partial_{[b} \Lambda_{a]}=0$.

The diffeomorphisms generated by $(\epsilon^a, \epsilon^i)$ with $\partial_b \epsilon^i=0$ are nothing but the foliation preserving diffeomorphisms, which we denote ${\rm Fdiff}(M,L)$.
They map a leaf to another leaf while preserving the foliation $L$.
Therefore, assuming a Dirac structure on a target space that admits a D-brane, 
breaks the symmetry from ${\rm Diff}(M)$ to ${\rm Fdiff}(M,L)$.
To preserve both $L$ and $L^*$ of the generalized product structure $L\oplus L^*$, 
another condition $\partial_i \epsilon^a=0$ is necessary.
Note that ${\rm Fdiff}(M,L)$ includes a global symmetry of the transverse displacement of leaves generated by 
$\epsilon^i={\rm const.}$
Specifying a particular leaf $L_p$ at $p\in M$ as a D-brane corresponds to 
the spontaneous symmetry breaking of this symmetry.
The scalar fields $\Phi^i$ are the corresponding NG-bosons.
In this case, ${\rm Diff}(M)$ is broken to a diffeomorphism preserving the leaf $L_p$, which we denote ${\rm LDiff}(M,L_p)$%
\footnote{It is not a subgroup of ${\rm Fdiff}(M,L)$ in general. 
There is another notion of a leaf preserving diffeomorphism, which is a subgroup of ${\rm Fdiff}(M,L)$ generated by $\epsilon^a$ ($\epsilon^i=0$) that maps each leaf to itself.
But it is too restrictive as a condition to keep a single leaf.},
and it is specified by a condition $\epsilon_\perp |_{L_p}=0$.

B-field gauge transformations are accompanied by this structure of diffeomorphisms.
${\rm Fdiff}(M,L)$ is paired with gauge transformations that are
 generated by $(\Lambda_i, \Lambda_a)$ with $\partial_{[b} \Lambda_{a]}=0$,
or equivalently $d_L \Lambda_\parallel =0$.
Note that $\Lambda_\parallel \in L^*$ so that $d_L \Lambda_\parallel \in \wedge^2 L^*$.
Therefore, this condition says that the deformation of $L$ by a tensor 
 $d_L \Lambda_\parallel$ does not change $L$, i.e., $L+(d_L \Lambda_\parallel)(L)=L$. 
(On the other hand, $L_{\cal F}=L+{\cal F}(L)$ is a true deformation.)
In particular, the transformation by a constant $\Lambda_{a}$ is a global gauge transformation, 
and a gauge field $A_a$ 
can be considered as 
 a NG-boson corresponding to this symmetry.
This will become more apparent after deriving the non-linear transformation 
laws for scalar and gauge fields in the next section.
Note also that in the presence of the gauge field, hidden transformations by exact 1-forms 
$\Lambda =d\lambda$ become visible as a $U(1)$ gauge symmetry.

\subsection{Non-linear transformation law}
\label{subsec:gene_Liederiv:4kfg}

Here, we derive a non-linear transformation law for scalar and gauge fields $\Phi+A$ under the diffeomorphism and the B-field gauge transformations.
The strategy is as follows.
In general, ${\rm Diff}(M)\ltimes \Omega^2_{\rm closed}(M)$ is a Courant automorphism and  
it is guaranteed that a Dirac structure is mapped to another Dirac structure.
On the other hand, any Dirac structure can be written as a graph such as $L_{\cal F}=L+{\cal F}(L)$.
Combining these, a Dirac structure $L_{\cal F}$ is mapped to another Dirac structure
$L_{\cal F'}$.
This determines a transformation law for the tensor ${\cal F}$.

\begin{figure}[tbp]
\begin{center}
\begin{minipage}{12cm}
\begin{minipage}{0.5\hsize}
\begin{center}
\includegraphics[scale=0.45]{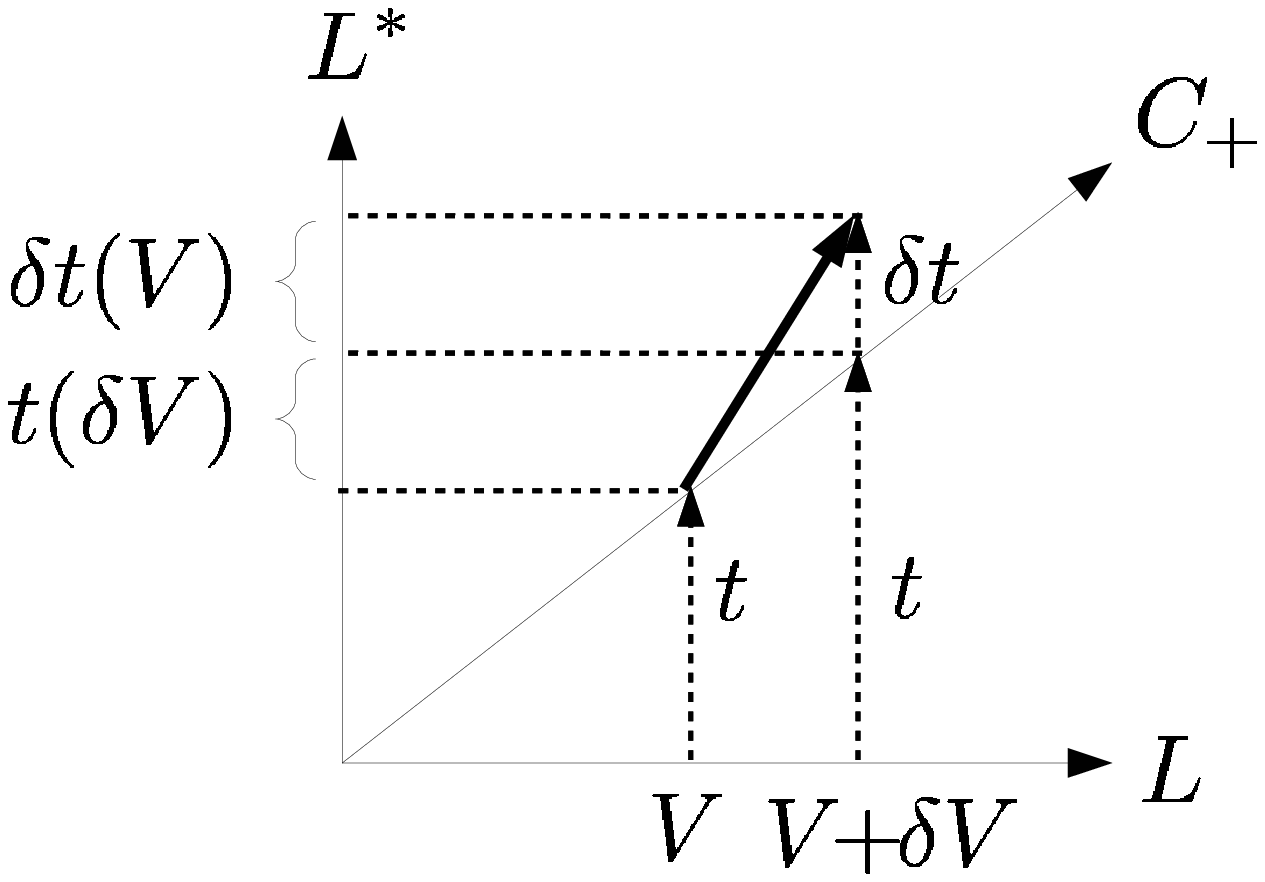}
\end{center}
\end{minipage}
\begin{minipage}{0.5\hsize}
\begin{center}
\includegraphics[scale=0.45]{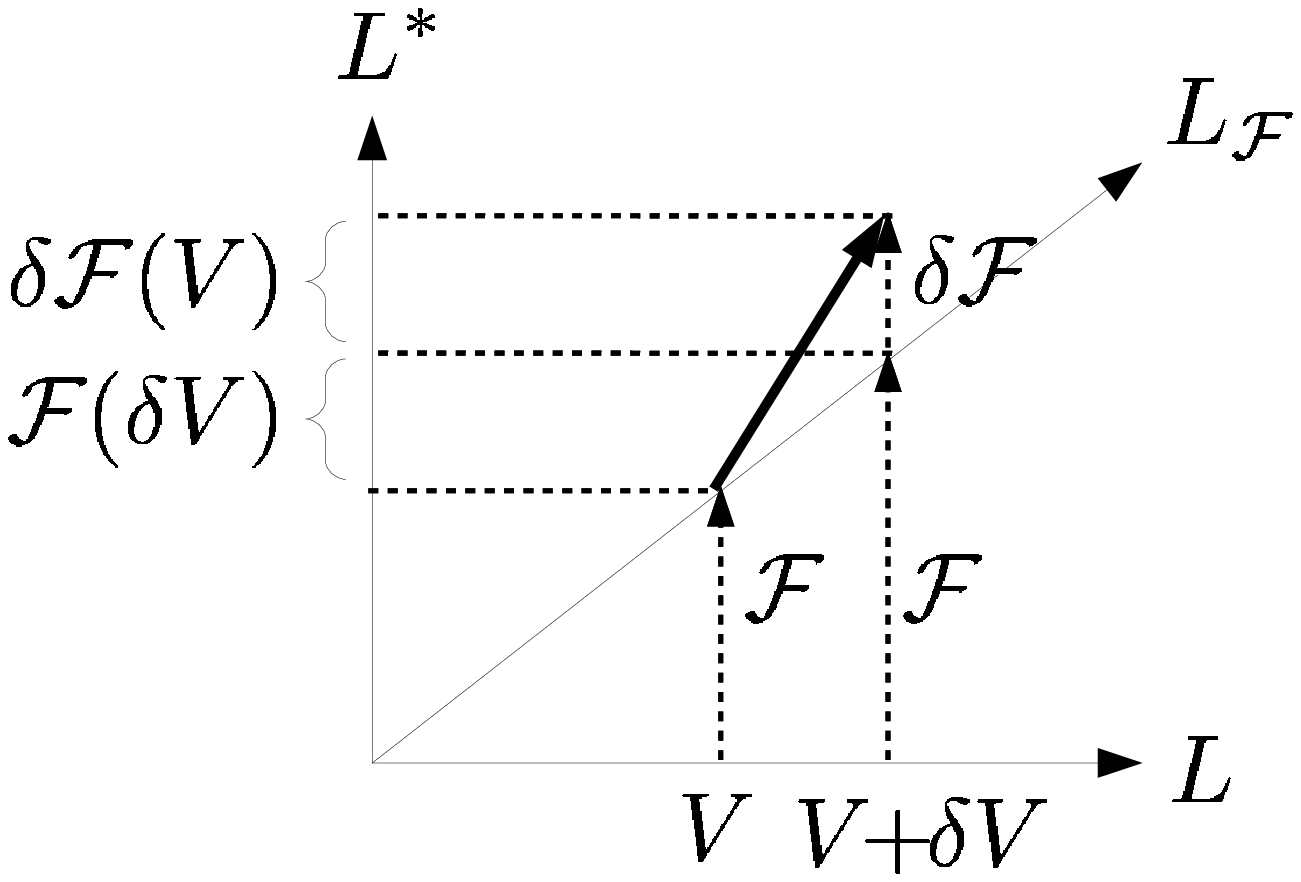}
\end{center}
\end{minipage}
\caption{\footnotesize The definition of the non-linear transformation  $\delta T$
 of the tensor $T$ defining the $C_+$, and  $\delta\cal F$ 
of the tensor $\cal F$ defining $L_{\cal F}$. They are defined by the deviation from a section given by
 the image of $V+\delta V\in L$.  }
\label{fig:gene_Lie_derivative}
\end{minipage}
\end{center}
\end{figure}

We derive the formula in a slightly more general context, 
since in the next section it is also used when we consider a generalized metric.
See also the appendix \ref{subsec:derivation} for details.
Let $T \in \Gamma(L^* \otimes L^*)$ be a tensor and define 
a graph $L_T=L+T(L)$, which need not be a Dirac structure.
On a section $V+T(V) \in L_T$ with $V\in L$, consider an action 
of a generalized Lie derivative ${\mathcal L}_{\epsilon+\Lambda}$ 
and rewrite it in the form (See Fig.\ref{fig:gene_Lie_derivative})
\begin{equation}
\begin{split}
-{\mathcal L}_{\epsilon+\Lambda} (V+T(V)) &= 
-{\mathcal L}_{\epsilon+\Lambda} V - {\mathcal L}_{\epsilon+\Lambda}(T(V))\\
& \overset{\rm def.}{=} \delta V + T(\delta V) + (\delta T)(V).
\end{split} \label{eq:def_liederoffield:3ld0}
\end{equation}
Here $\delta V \in L$ denotes the horizontal shift determined by projecting the transformed 
section to the $L$-direction. Thus, the first two terms in (\ref{eq:def_liederoffield:3ld0}) 
represent a shift along $L_T$, 
and the remaining part denoted by $\delta T \in \Gamma(L^* \otimes L^*)$ 
 gives a net deformation of the graph $L_T$.
The combination $T'=T+\delta T$ is the desired tensor giving the transformed graph $L_{T'}$.

Note that for a Dirac structure $L_{\cal F}$, we can also derive the same formula
by noting that $L_{\cal F}=e^{-{\cal L}_{\Phi+A}}L$.
In this case, the calculation reduces to a commutator 
$[{\cal L}_{\epsilon +\Lambda}, {\cal L}_{\Phi+A}]$. 

From the result (\ref{eq:Lie_field_LL:4e3da}), we read off in the case of our Dirac structure 
$L_{\cal F}$:\begin{subequations}
\begin{align}
\delta{\mathcal F}_{ab} &= -\epsilon^M \partial_M {\mathcal F}_{ab} -\partial_a \epsilon^c {\mathcal F}_{cb} - {\mathcal F}_{ac}\partial_b \epsilon^c -\partial_{[a} \Lambda_{k]} {\mathcal F}^k_{\ b} +{\mathcal F}_a^{\ k}\partial_{[k} \Lambda_{b]} \\
&\quad -{\mathcal F}_a^{\ k}\partial_k \epsilon^c {\mathcal F}_{cb}+{\mathcal F}_{ac}\partial_k\epsilon^c {\mathcal F}^k_{\ b} -{\mathcal F}_a^{\ k}\partial_{[k}\Lambda_{l]} {\mathcal F}^{l}_{\ b} +\partial_{[a}\Lambda_{b]}, \notag \\
\delta{\mathcal F}_a^{\ j} &=-\delta{\mathcal F}^j_{\ a}= -\epsilon^M \partial_M {\mathcal F}_a^{\ j} -\partial_a \epsilon^c {\mathcal F}_c^{\ j} +{\mathcal F}_a^{\ k}\partial_k \epsilon^j -{\mathcal F}_a^{\ k}\partial_k \epsilon^c {\mathcal F}_{c}^{\ j} +\partial_a\epsilon^j ,
\end{align} \label{eq:gene_lie_abelianfield:4fl2s}
\end{subequations}
or equivalently,
\begin{subequations}
\begin{align}
\delta{\mathcal F}_{ab} &= (\partial_{[a}+ {\mathcal F}_{[a}^{\ \ j} \partial_j) (\Lambda_{b]}-\epsilon^c {\mathcal F}_{cb]}-\Lambda_k {\mathcal F}^k_{\ b]}) - (\epsilon^k - \epsilon^c {\mathcal F}_c^{\ k}) \partial_k {\mathcal F}_{ab}, \\
\delta{\mathcal F}_a^{\ j} &= (\partial_{a}+ {\mathcal F}_{a}^{\ \ i} \partial_i) (\epsilon^j - \epsilon^c {\mathcal F}_c^{\ j}) - (\epsilon^k - \epsilon^c {\mathcal F}_c^{\ k}) \partial_k {\mathcal F}_a^{\ j}.
\end{align}\label{eq:gene_Liederi_F:e47h}
\end{subequations}
When we evaluate them on the leaf of $x^i=\Phi^i$, using (\ref{eq:gene_fieldstrength_abelian:3kf9}),
they correspond to the transformation law for a gauge field 
and scalar fields as 
\begin{subequations}
\begin{align}
\delta A_a &= \Lambda_a-\epsilon^c F_{ca}+\Lambda_k\partial_a\Phi^k, \\
\delta \Phi^i &= \epsilon^i - \epsilon^c \partial_c \Phi^i,
\end{align} \label{eq:gene_Liederi_Aphi:1g6j}
\end{subequations}
Here the terms including $\epsilon^c$ and $\Lambda_k$ represent unbroken symmetry, 
the ordinary diffeomorphism along leaves (worldvolume) 
and the unbroken B-field gauge transformation induced on a leaf
\footnote{The term $-\epsilon^c F_{ca}$ is rewritten as a ordinary tensor transformation by using a $U(1)$ gauge transformation $A_a \to A_a-\partial_a (\epsilon^c A_c)$.}.
These transformations of unbroken symmetry are 
linear in the fields $A_a$ and $\Phi^i$, as desired.
On the other hand, the first term in each line is inhomogeneous in the 
fields, and it corresponds to the generators of broken symmetry.
As we have discussed, the scalar fields $\Phi^i$ 
is a NG-boson for the global displacement,
and the formula above is exactly the non-linearly realized 
transformation law for a NG-boson.
In the same way, we conclude that a gauge field is a NG-boson 
for a broken global B-field gauge transformation.

Note that these transformation laws are the extension of 
non-linearly realized Poincar\'e symmetry in \cite{Gliozzi:2011hj,Casalbuoni:2011fq}.
They derived the formula in the context of a field theory 
on a worldvolume $\Sigma$ and a compensating diffeomorphism.
We reproduce and extend their result in purely geometric terms.
In particular, it becomes possible to treat a gauge field in a same way 
as scalar fields, since a gauge field is also cast into a geometry.

\section{Adding generalized metric}
\label{sec:gene_metric:p3ke}

In the previous section, we analyzed the geometrical structure 
which is associated with a D-brane and the open strings from the
physical point of view.
Independently, there is a generalized Riemannian structure coming 
from the closed strings, as reviewed in \S 2.
In this section, we consider both structures. 

\subsection{Generalized metric seen by a D-brane}
\label{subsec:gene_seen_by_L}

Recall that a generalized metric is given by a graph $C_+=\{X+(g+B)(X)\,|\,X\in TM\}$.
The main observation here is that the same argument is applied not only $TM$ but also any other Dirac structure $L$.
This is possible because its dual Dirac structure $L^\ast$ is also isotropic and the intersection $L^\ast \cap  C_+$ is the zero section only.
Thus, the generalized Riemannian structure is always written by a graph $C_+=\{V+t(V)\,|\,V\in L\}$ of some appropriate tensor $t\in \wedge^2 L^*$ viewed as a map $t: L \to L^\ast $.

In our case, given a Dirac structure $L$, a section of $C_+$ is thus written as
\begin{equation}
V_+ = v^a(\partial_a + t_a^{\ j}\partial_j + t_{ab} dx^b) + \xi_i (dx^i + t^i_{\ b}dx^b + t^{ij}\partial_j) \quad \in \Gamma(C_+).
\label{eq:VL0_genemet_0e3da}
\end{equation}
Since (\ref{eq:VL0_genemet_0e3da}) and  (\ref{eq:Genemetricasgraph:3kdo}) are the expression of the same element in $C_+$, they should be identified.
This determines the tensor $t \in \wedge^2 L^*$, and 
we find the relations (for a proof, see \S\ref{subsec:app_genemet_as_graphs}) 
\begin{equation}
\begin{split}
t^{ij} &= E^{ij}, \qquad \qquad t_a^{\ j}= - E_{ak}E^{kj}, \\
t^i_{\ b} &= E^{ik}E_{kb}, \qquad \  t_{ab} = E_{ab} - E_{ak} E^{kl} E_{lb}.
\label{eq:D-brane_Buscher_rule:4kos}
\end{split}
\end{equation}
where $E^{ji}$ is the inverse of $E_{ji}$, satisfying $E^{kj}E_{ji}=\delta^k_i$. 

These relations (\ref{eq:D-brane_Buscher_rule:4kos}) are similar to the Buscher rule \cite{Buscher:1987sk,Buscher:1987qj}, i.e., the T-duality rule for a background field $E=g+B$. 
However, this is not exactly a T-duality, since
 a T-duality is in general a map between two different spacetimes, and is an operation which exchanges 
a basis of vector fields and a basis of 1-forms \cite{Grana:2008yw,Cavalcanti:2011wu,Bouwknegt:2003zg}. This means that the rule of exchanging upper 
and lower indices lies at the heart of the T-duality rule%
\footnote{The lowering of 
the indices $i$ and $j$ in (\ref{eq:D-brane_Buscher_rule:4kos}) to form a tensor $t_{MN}$ 
(i.e., $t \in \Gamma(\wedge^2 T^*\tilde{M})$ for the T-dual manifold $\tilde{M}$).
}.
On the other hand, (\ref{eq:D-brane_Buscher_rule:4kos}) shows that the mixing of a metric $g$ and a B-field $B$ is already seen without T-duality.

This observation sheds some light on the question raised in \S 2.
There are at least three kinds of symmetric tensors $g$, $g-Bg^{-1}B$ and $g^{-1}$ 
that appear in a generalized Riemannian structure.
The first one is a positive definite metric defined on a graph $u+(g+B)(u) \in C_+$, which originates from closed strings. 
The others appear in the off-diagonal elements of $G$ in (\ref{eq:indgenemetG_oscj9d}).
Here $g-Bg^{-1}B$ is the restriction of $G$ to $TM$, but $TM$ is an example of a Dirac structure, 
corresponding to a D$9$-brane in the case of $10$-dimensional spacetime $M={\mathbb R}^{10}$.
Similarly,  $g^{-1}$ is the restriction to $T^*M$, which is the 
Dirac structure corresponding to a D-instanton.
This suggests that the latter two symmetric 
tensors are closely related to open strings\footnote{It is interesting that $g-Bg^{-1}B$ is the symmetric part of $(g+B)^{-1}$, 
which is already referred to as the ``open string metric" in the literature.}.

Coming back to our situation, let us write an endomorphism $G$ 
in (\ref{eq:indgenemetG_oscj9d}) 
as a matrix in the basis of $L\oplus L^*$.  
By solving the eigenvalue equation (\ref{eq:eigenvalueeq:8hst}), we get
\begin{equation}
G = \begin{pmatrix} -s^{-1}a & s^{-1} \\ s-as^{-1}a & as^{-1} \end{pmatrix},
\end{equation}
where $s$ and $a$ are symmetric and anti-symmetric parts of the tensor $t=s+a$. 
By restricting it to $L$, we obtain $s-as^{-1}a$, 
which is a candidate of the metric seen by a D-brane without fluctuations.

The inclusion of the fluctuations is straightforward in this picture by using the generalized field strength ${\cal F}$.
Suppose that the generalized metric $C_+$ is seen by a Dirac structure $L_{\cal F}=L+{\cal F}(L)$ through a tensor $t_{\cal F} \in L^*\otimes L^*$.
That is, $C_+=L_{\cal F}+t_{\cal F}(L_{\cal F})$.
Then a section of $C_+$ is written for $V\in L$,
\begin{align}
V+{\cal F}(V)+t_{\cal F}(V+{\cal F}(V))
=V+(t_{\cal F}+{\cal F})(V),\label{eq:CplusfluctuatedVector}
\end{align}
where we have used the fact that $t_{\cal F}{\cal F}(V)=0$ since ${\cal F}(V)\in L^*$.
The vector in eq.(\ref{eq:CplusfluctuatedVector}) should be identified with $V+t(V)$, thus 
we have $t_{\cal F}=t-{\cal F}$.
Therefore, a generalized field strength on a D-brane always appears 
as a shift of the tensor $t$ when considering a generalized Riemannian structure.
This is also true when we write $G$ as a matrix 
of a map $L_{\cal F}\oplus L_{\cal F}^\ast \to L_{\cal F}\oplus L_{\cal F}^\ast$, that is,
\begin{equation}
G = \begin{pmatrix} -s^{-1}(a-{\mathcal F}) & s^{-1} \\ s-(a-{\mathcal F})s^{-1}(a-{\mathcal F}) & (a-{\mathcal F})s^{-1} \end{pmatrix}.
\end{equation}
By restriction of $G$ to $L_{\mathcal F}$, the metric on $L_{\mathcal F}$ is given by
\begin{equation}
s_{\mathcal F}\overset{\rm def.}{=} s-(a-{\mathcal F})s^{-1}(a-{\mathcal F}) \in \Gamma(L^* \otimes L^*),
\label{def s_F}
\end{equation}
where the curvature ${\cal F}$ appears as a shift of the anti-symmetric part $a \in \wedge^2 L^*$ of the generalized Riemannian structure.
This fact plays a central role when considering the DBI action 
in \S 6.

\subsection{Symmetry transformations of generalized metric}
\label{sec:genelie_metric:e3lf}

Before proceeding the discussion, we study the effect of the symmetry ${\rm Diff}(M)\ltimes \Omega^2_{\rm closed}(M)$ on a generalized Riemannian structure $C_+ \subset {\mathbb T}M$.

Recall that a metric $g$ and a B-field $B$ are dynamical fields 
of a rank $2$ tensor in the target space $M$.
That is, an infinitesimal transformation of $E=g+B$ is given 
as $\delta E=-{\cal L}_{\epsilon +\Lambda}E$ 
by an action of a generalized Lie derivative, 
in components, as
\begin{equation}
\delta E_{MN} = - \epsilon^L \partial_L E_{MN} - \partial_M \epsilon^LE_{LN} - E_{ML}\partial_N\epsilon^L + \partial_{[M}\Lambda_{N]}.
\label{tensor rule}
\end{equation}

On the other hand, a generalized metric $C_+$ is also defined as a graph of $t : L\to L^*$.
Then, an action of a generalized Lie derivative ${\cal L}_{\epsilon +\Lambda}$ on $C_+$ reduces to a non-linear transformation for the tensor $t\in \Gamma(L^* \otimes L^*)$ as argued in \S \ref{subsec:gene_Liederiv:4kfg}. 
The result is (replace $T$ with $t$ given in (\ref{eq:Lie_field_LL:4e3da}))
\begin{subequations}
\begin{align}
\delta t_{ab} &= -\epsilon^M \partial_M t_{ab} -\partial_a \epsilon^c t_{cb} - t_{ac}\partial_b \epsilon^c +\partial_{[k} \Lambda_{a]} t^k_{\ b} +t_a^{\ k}\partial_{[k} \Lambda_{b]} \notag\\
& \quad -t_a^{\ k}\partial_k \epsilon^c t_{cb}+t_{ac}\partial_k\epsilon^c t^k_{\ b} -t_a^{\ k}\partial_{[k}\Lambda_{l]} t^{l}_{\ b} +\partial_{[a}\Lambda_{b]}, \\
\delta t_a^{\ j} &= -\epsilon^M\partial_M t_a^{\ j} -\partial_a \epsilon^c t_c^{\ j} +t_a^{\ k}\partial_k \epsilon^j  +\partial_{[k}\Lambda_{a]} t^{kj} \notag \\
& \quad -t_a^{\ k}\partial_k \epsilon^c t_{c}^{\ j} +t_{ac}\partial_k\epsilon^c t^{kj} -t_a^{\ k}\partial_{[k}\Lambda_{l]} t^{lj} +\partial_a \epsilon^j, \\
\delta t^i_{\ b} &= -\epsilon^M\partial_M t^i_{\ b} +\partial_k \epsilon^i t^k_{\ b} -t^i_{\ c}\partial_b \epsilon^c  + t^{ik}\partial_{[k}\Lambda_{b]} \notag \\
& \quad -t^{ik}\partial_k \epsilon^c t_{cb} +t^i_{\ c}\partial_k\epsilon^c t^k_{\ b} -t^{ik}\partial_{[k}\Lambda_{l]}t^{l}_{\ b} -\partial_b \epsilon^i, \\
\delta t^{ij} &=-\epsilon^M \partial_M t^{ij} +\partial_k \epsilon^i t^{kj} +t^{ik}\partial_k \epsilon^j \notag\\
& \quad -t^{ik} \partial_k \epsilon^c t_c^{\ j} +t^i_{\ c} \partial_k \epsilon^c t^{kj} -t^{ik} \partial_{[k}\Lambda_{l]}t^{lj}.
\end{align} \label{eq:Lie_of_genemet_tensor:ek4l}
\end{subequations}
At first sight, it is unexpected that such a non-linear law appears.
However, it is consistent with the tensor rule (\ref{tensor rule}) above.
In fact, the following diagram is shown to be commutative (see appendix \ref{subsec:consistency} for detail):
\begin{equation}
\xymatrix{ \ar@{} [rd] E \ar[r]^{\epsilon+\Lambda~~~~} \ar[d] 
& E+\delta E \ar[d] \\
t \ar[r]_{\epsilon+\Lambda~~~~} & t+\delta t}\label{commutativeDG}
\end{equation}
Here the vertical arrow represents the map given by the relation (\ref{eq:D-brane_Buscher_rule:4kos}).
The appearance of the inhomogeneous terms in $\delta t$ is simply due to the fixing of a Dirac structure $L$ as a reference frame, where the transformation of $L$ itself is absorbed into that of a tensor $t$.
Therefore, such a inhomogeneous law is a general feature valid for any tensor in a target space.

By combining two results (\ref{eq:gene_lie_abelianfield:4fl2s}) and (\ref{eq:Lie_of_genemet_tensor:ek4l}), the transformation law for a combination $t_{\mathcal F}=t-{\mathcal F}$ is obtained:
\begin{subequations}
\begin{align}
\delta {t_{\mathcal F}}_{ab} &= -\epsilon^M \partial_M {t_{\mathcal F}}_{ab} -\partial_a \epsilon^c {t_{\mathcal F}}_{cb} - {t_{\mathcal F}}_{ac}\partial_b \epsilon^c +\partial_{[k} \Lambda_{a]} {t_{\mathcal F}}^k_{\ b} +{t_{\mathcal F}}_a^{\ k}\partial_{[k} \Lambda_{b]} \notag\\
& \quad -t_a^{\ k}\partial_k \epsilon^c t_{cb}+t_{ac}\partial_k\epsilon^c t^k_{\ b} -t_a^{\ k}\partial_{[k}\Lambda_{l]} t^{l}_{\ b} \\
& \quad +{\mathcal F}_a^{\ k}\partial_k \epsilon^c {\mathcal F}_{cb}-{\mathcal F}_{ac}\partial_k\epsilon^c {\mathcal F}^k_{\ b} +{\mathcal F}_a^{\ k}\partial_{[k}\Lambda_{l]} {\mathcal F}^{l}_{\ b} \notag, \\
\delta {t_{\mathcal F}}_a^{\ j} &= -\epsilon^M\partial_M {t_{\mathcal F}}_a^{\ j} -\partial_a \epsilon^c {t_{\mathcal F}}_c^{\ j} +{t_{\mathcal F}}_a^{\ k}\partial_k \epsilon^j  +\partial_{[k}\Lambda_{a]} {t_{\mathcal F}}^{kj} \notag \\
& \quad -t_a^{\ k}\partial_k \epsilon^c t_{c}^{\ j} +t_{ac}\partial_k\epsilon^c t^{kj} -t_a^{\ k}\partial_{[k}\Lambda_{l]} t^{lj} +{\mathcal F}_a^{\ k}\partial_k \epsilon^c {\mathcal F}_{c}^{\ j}, \\
\delta {t_{\mathcal F}}^i_{\ b} &= -\epsilon^M\partial_M {t_{\mathcal F}}^i_{\ b} +\partial_k \epsilon^i {t_{\mathcal F}}^k_{\ b} -{t_{\mathcal F}}^i_{\ c}\partial_b \epsilon^c  + {t_{\mathcal F}}^{ik}\partial_{[k}\Lambda_{b]} \notag \\
& \quad -t^{ik}\partial_k \epsilon^c t_{cb} +t^i_{\ c}\partial_k\epsilon^c t^k_{\ b} -t^{ik}\partial_{[k}\Lambda_{l]}t^{l}_{\ b} +{\mathcal F}^{ik}\partial_k \epsilon^c {\mathcal F}_{cb}, \\
\delta {t_{\mathcal F}}^{ij} &=-\epsilon^M \partial_M {t_{\mathcal F}}^{ij} +\partial_k \epsilon^i {t_{\mathcal F}}^{kj} +{t_{\mathcal F}}^{ik}\partial_k \epsilon^j \notag\\
& \quad -t^{ik} \partial_k \epsilon^c t_c^{\ j} +t^i_{\ c} \partial_k \epsilon^c t^{kj} -t^{ik} \partial_{[k}\Lambda_{l]}t^{lj},
\end{align}
\end{subequations}
Remarkably, inhomogeneous terms $\partial_b \epsilon^i$ 
and $\partial_{[b}\Lambda_{a]}$ 
in (\ref{eq:gene_lie_abelianfield:4fl2s}) and (\ref{eq:Lie_of_genemet_tensor:ek4l}) 
cancel each other, and do not appear in the combination $t_{\mathcal F}=t-{\mathcal F}$.
This is because $t_{\cal F}$ represents a difference between $C_+$ and $L_{\cal F}$, 
which is independent of the fixed frame $L$.
This will be ultimately related to a similar kind of cancellation found in \cite{Gliozzi:2011hj} for the case of a non-linearly realized Poincar\'e transformation, 
where it comes only from ${\cal F}$.
Here we emphasize that the appearance of a combination $t_{\mathcal F}=t-{\mathcal F}$ and a cancellation of inhomogeneous terms have geometric explanations.


\section{Dirac-Born-Infeld action}
\label{sec:non-linear realization}
So far we have studied the characterization and symmetry of D-brane and its fluctuations 
form a target space viewpoint in
the framework of the generalized geometry.
In this section, we focus on a field theory on a D-brane worldvolume $\Sigma$.
In particular, we study the DBI action.
First, we give a simple proof of the invariance of the DBI 
action under the non-linear realized spacetime symmetries.
In the successive section, we show how to construct the DBI action merely from the knowledge of the
geometrical information and the symmetry without referring to the string.

\subsection{Invariance of Dirac-Born-Infeld Action}\label{sec:inv_DBI}

First we give here the proof that the DBI action has a symmetry 
under the transformation given in (\ref{eq:gene_Liederi_Aphi:1g6j})
of full ${\rm Diff}(M)\ltimes \Omega^2_{\rm closed}(M)$
transformation. 
This can be proven by representing the 
DBI action as a product of
the determinants of the two types of metrics 
which we found in the analysis of the generalized Riemannian structure in the previous section.
As we see, their transformation under the generalized Lie derivative has
a rather simple form.

We first prove the following relation:
\begin{align}
\begin{split}
{\det}^\frac{1}{4} g \, {\det}^\frac{1}{4} s_{\cal F}
=& \sqrt{\det (\varphi_\Phi^\ast (g+B) -F )_{ab}}.
\label{Lag_DBI}
\end{split}
\end{align}
where $g$ is the Riemannian metric on $TM$, i.e., $g\in T^*M\otimes T^*M$ and 
$s_{\cal F}$ is the metric on $L_{\cal F}$, i.e., $s_{\cal F}\in L_{\cal F}^*\otimes L_{\cal F}^*$
given in (\ref{def s_F}).
$\varphi_\Phi^*$ is the pullback of the embedding map defined by the field $\Phi$, 
$\varphi_\Phi: \Sigma \hookrightarrow M$%
\footnote{Precisely, $\varphi_\Phi^*$ here denotes the tensor structure of the pull-back 
as $\varphi^*_\Phi(E)_{ab}=E_{ab}+\partial_a\Phi^i E_{ib}+ E_{aj}\partial_b\Phi^j +\partial_a\Phi^i\partial_b\Phi^j E_{ij}$
and it is defined not only on $\Sigma$ but on whole $M$.
}.
Thus determinants on the
 l.h.s. are those of the $D\times D$ matrices,
while the determinant on the r.h.s. is that of the $(p+1)\times (p+1)$ matrix which is distinguished by the
index $ab$.
The r.h.s. of (\ref{Lag_DBI}) is the Lagrangian ${\cal L}_{\rm DBI}$ of the DBI action.

This equation can be proven by combining the following relations of various determinants.
\begin{enumerate}
\item Let $s$ be the symmetric part of $t$ defined in eq.(\ref{eq:D-brane_Buscher_rule:4kos}) and $t^{ij}=E^{ij}$ the inverse matrix of $E_{ij}$,
then
\begin{equation}
\det s=\det g(\det t^{ij})^2~.\label{detgt2}
\end{equation} 
\item From the definition (\ref{def s_F}) of $s_{\mathcal F}$, 
\begin{equation}
\det s\det s_{\cal F}=(\det t_{\cal F})^2~.\label{dettF2}
\end{equation}
\item Using the explicit expression for $t_{\cal F}$,
\begin{equation}
\det t_{\cal F}=\det t^{ij}\det(\varphi_\Phi^\ast (g+B) -F )_{ab}~.\label{dettDBI}
\end{equation}
\end{enumerate}
The derivation of these relations are given in the appendix \ref{subsec:identity}. 
Using these relations
it is straightforward to prove the representation of the DBI action given in eq.(\ref{Lag_DBI}):
\begin{align}
\begin{split}
{\det}^\frac{1}{4} g \, {\det}^\frac{1}{4} s_{\cal F}
=&  \frac{1}{\sqrt{\det t^{ij}}}\,{\det}^\frac{1}{4} s \,{\det}^\frac{1}{4} s_{\cal F} \\
=& \frac{1}{\sqrt{\det t^{ij}}}\, \sqrt{\det t_{\cal F}} \\
=& \sqrt{\det (\varphi_\Phi^\ast (E) -F )_{ab}}.
\label{L_DBIproof}
\end{split}
\end{align}
Here we have used (\ref{detgt2}) in the first equality, (\ref{dettF2}) in the second equality,
and (\ref{dettDBI}) for the last step.

The integral of this scalar density (\ref{Lag_DBI}) agrees with the DBI action
\begin{equation}
S_{\rm DBI} = \int_{\varphi_\Phi (\Sigma)} \sqrt{\det (\varphi_\Phi^\ast (g+B) -F )_{ab}} \ dx^0 \wedge \cdots \wedge dx^p,
\label{DBIaction}
\end{equation}
when evaluated on the leaf $\varphi_\Phi (\Sigma)$ of $L_{\cal F}$ at $x^i=\Phi^i (x)$.
Apparently, the DBI action is invariant under the worldvolume diffeomorphism on the D-brane. 

Now we can prove that the DBI action is not only invariant under the world volume diffeomorphism but also under the full target space diffeomorphism and the B-field gauge transformation.
To this end, we rewrite (\ref{DBIaction}) as an integral over the target space $M$ as
\begin{equation}
S_{\rm DBI} = \int_M {\cal L}_{\rm DBI} \ \delta^{(D-p-1)}(x^i-\Phi^i(x^a)) \ dx^0 \wedge \cdots \wedge dx^{D-1},
\label{DBI2}
\end{equation}
where ${\cal L}_{\rm DBI}$ is given in (\ref{Lag_DBI}) and $\delta^{(D-p-1)}(x^i-\Phi^i(x^a))$ 
is a Dirac's delta function seen as a distribution along $x^i$-directions.

The infinitesimal transformation of the full diffeomorphism and the $B$-field gauge transformation are 
parametrized by $\epsilon +\Lambda $, 
and are studied in the previous section (also summarized in the appendix \ref{subsec:app_gene_lie_deri_various_str:3kfs}).
The transformation of the integrand ${\cal L}_{\rm DBI}$ is 
obtained from that of ${\det g}$ 
\begin{equation}
\delta \det g = -\epsilon^M \partial_M \det g +\det g\left\{ -2\partial_M\epsilon^M \right\} 
\label{transformation of det g}~,
\end{equation}
and that of $\det s_{\cal F}$ 
\begin{equation}
\delta \det s_{\mathcal F} 
= -\epsilon^M \partial_M \det s_{\mathcal F} +\det s_{\mathcal F} \left\{ -2\partial_c\epsilon^c +2\partial_k\epsilon^k -4{\mathcal F}_c^{\ k}\partial_k\epsilon^c \right\} \label{root s_F}~.
\end{equation}
The result is
\begin{equation}
\delta {\cal L}_{\rm DBI} 
= -\epsilon^M \partial_M {\cal L}_{\rm DBI} -(\partial_c \epsilon^c +\partial_c \Phi^k \partial_k \epsilon^c){\cal L}_{\rm DBI}.
\label{Lie derivative of DBI}
\end{equation}
Except for the first term, this depends only on $\epsilon^a$, i.e., 
other parameters $\epsilon^i$ and $\Lambda_M$ are absent.
This is the expected result since ${\cal L}_{\rm DBI}dx^0 \wedge \cdots \wedge dx^p$ 
is a section of $\det(\Delta^*)$ as we will explain in the next section.
On the other hand, the delta function transforms as
\begin{align}
&\delta [\delta^{(D-p-1)}(x^i-\Phi^i)]\nonumber\\
=& -\epsilon^M \partial_M [\delta^{(D-p-1)}(x^i-\Phi^i)]
-(\partial_k\epsilon^k-\partial_k\epsilon^c\partial_c\Phi^k) \delta^{(D-p-1)}(x^i-\Phi^i).
\label{Lie derivative of delta} 
\end{align}
By combining them, we obtain 
\begin{equation}
\delta\left[{\cal L}_{\rm DBI} \,\delta^{(D-p-1)}(x^i-\Phi^i)\right]
= -\partial_M \left[\epsilon^M{\cal L}_{\rm DBI} \, \delta^{(D-p-1)}(x^i-\Phi^i)\right] ~.
\end{equation}
Namely the transformation of the integrand in the DBI action (\ref{DBI2}) is a total derivative and 
 the DBI action is invariant under full target space diffeomorphisms and B-field gauge transformations.

This invariance itself is in some sense trivial, because a generalized Riemannian structure $g+B$ is also transformed.
Since a shape of a leaf and a metric on a leaf is changed simultaneously, its volume is unchanged. 
We stress that the non-trivial thing here is the invariance within the static gauge.

\subsection{Non-linear symmetry and effective action}
\label{sec:inv_Effective}

In this section, we analyze how much the non-linearly realized symmetry restricts the 
form of the effective action.

We would like to find possible ingredients to build an effective action for a D-brane,
that is made out of fields $A+\Phi$ on the worldvolume, as well as a generalized metric $g+B$.
Usually, an action is an integral over the worldvolume $\Sigma$, 
and its integrand is a scalar density under the diffeomorphism on the worldvolume.
In our setting, where a reference generalized product structure is fixed (static gauge), 
such a worldvolume object is given by restricting a target space object to a leaf, as seen in the following.

Recall that a scalar density in a target space $M$ can be considered as a section of the determinant bundle ${\det}(T^*M)$ over $M$ associated with the cotangent bundle $T^*M$.
It is equivalently seen as top forms $\wedge^{\rm top}T^*M$.
For example, a Riemannian%
\footnote{$\sqrt{-\det g}$ for a Lorentzian signature.}
 metric on $TM$, $g: TM\to T^*M$ (or equivalently a tensor $g \in T^*M \otimes T^*M$) defines a volume form $\sqrt{g} := \sqrt{\det g} dx^{0}\wedge \cdots \wedge dx^{D-1}$ as a section of  ${\det}(T^*M)$.
Note that $\det g$ denotes the determinant of a $D\times D$ matrix $g_{MN}$.
The volume form $\sqrt{g}$ transforms as a scalar density under ${\rm Diff}(M)$ by construction, 
and is invariant under the B-field gauge transformation.
In fact, by acting a generalized Lie derivative $-{\cal L}_{\epsilon +\Lambda }$ on $\sqrt{g}$, we obtain a tensor transformation law for its coefficient as 
\begin{equation}
\delta\sqrt{\det g}=\partial_M(\epsilon^M\sqrt{\det g}\label{root g})
\end{equation}
while the base $dx^0 \wedge \cdots \wedge dx^{D-1}$ is kept unchanged.%
This relation is used in (\ref{transformation of det g}).
Note that there is another choice of a section $\sqrt{E} = \sqrt{\det E} dx^{0}\wedge \cdots \wedge dx^{D-1}\in \det(T^* M)$ defined by a tensor $E=g+B$.
It is also a scalar density but is not invariant under the B-field gauge transformation.

Similarly, associated with Dirac structures $L$ or $L_{\cal F}$, any tensor $T\in L^* \otimes L^*$ defines a section of the determinant bundle $\det (L^*)$ of the form $\sqrt{T}:= \sqrt{\det T} dx^{0}\wedge \cdots \wedge dx^{p}\wedge \partial_{p+1}\wedge \cdots \wedge \partial_{D-1}$.
Note that $\det T$ is still a determinant of a $D\times D$ matrix, but with the upper index for $i$ and $j$: 
\begin{align}
\left(
\begin{array}{cc}
T_{ab} &T_a{}^j \\ T^i{}_b&T^{ij}
\end{array}\right).
\end{align}
The examples of the sections of $\det(L^*)$ can be constructed from the tensors that we have encountered
in the previous section. 
They are $\sqrt{t}$, $\sqrt{s}$, $\sqrt{s-as^{-1}a}$, $\sqrt{t_{\cal F}}$ and $\sqrt{ s_{\cal F}}$.
Their transformation law is deduced from the non-linear transformation law studied in \S 5.
Note that possible elements in $\det(T^* M)$ or $\det(L^*)$ are not independent but related through
\begin{align}
\det E &=\sqrt{\det g} \, \sqrt{\det (g-Bg^{-1}B)},\nonumber\\
\det t &=\sqrt{\det s} \, \sqrt{\det (s-as^{-1}a)},\nonumber\\
\det t_{\cal F} &=\sqrt{\det s} \, \sqrt{\det s_{\cal F}},
\label{symmetric decomposition}
\end{align}
which are easily shown (see appendix \ref{subsec:identity}) and used in (\ref{dettF2}).
It is interesting in its own right that the l.h.s. of eq.(\ref{symmetric decomposition}) 
is bulk (closed string) metric, while the r.h.s. 
is a product of open string metrics (see comments in \S 5).

On the other hand, a scalar density in the worldvolume $\Sigma$ is a section of ${\det}(T^*\Sigma)$.
Since $\Sigma$ is a leaf of a foliation $\Delta \subset TM$ (that is $\Delta|_\Sigma=T\Sigma$), 
we need an element of ${\det}(\Delta^* )$ as a scalar density on leaves.
To this end, it is useful to decompose sections of $\det(T^* M)$ or $\det(L^*)$ into a product of two sub-determinants.
More precisely, under the generalized product structure ${\mathbb T}M=L\oplus L^*$ with  $L=\Delta \oplus {\rm Ann}(\Delta)$, $T^*M$ and $L^*$ are split as $T^*M=\Delta^* \oplus {\rm Ann}(\Delta)$ and $L^*=\Delta^* \oplus {\rm Ann}^*(\Delta)$.
Then one can write the corresponding determinant bundles as
\begin{align}
{\det}(T^*M) &= \det (\Delta^* ) \otimes \det({\rm Ann}(\Delta)) ,\nonumber\\
\det (L^*) &=\det (\Delta^*) \otimes \det ({\rm Ann}^*(\Delta)) = \det (\Delta^*) \otimes {\det}^{-1} ({\rm Ann}(\Delta)).
\label{splitting}
\end{align}
The former corresponds to (the square root of) an identity on determinants
\begin{align}
\det \left(
\begin{array}{cc}
A_{ab} &B_{aj} \\ C_{ib}&D_{ij}
\end{array}\right)
=\det (A -BD^{-1}C)_{ab} \det D_{ij},
\label{submatrix identity}
\end{align}
valid for any $D\times D$ matrix with an invertible submatrix $D$,
where the two determinants on the r.h.s. are that those for $(p+1)\times (p+1)$ and 
$(D-1-p)\times(D-1-p)$ submatrices, respectively.
For the latter, indices $i$ and $j$ should be raised.
We give two examples associated to the splittings of the determinant bundle in (\ref{splitting}):
\begin{align}
& \sqrt{\det E}=\sqrt{\det t_{ab}}\sqrt{\det E_{ij}},\nonumber\\
& \sqrt{\det t}=\sqrt{\det E_{ab}}\sqrt{\det t^{ij}},
\label{decomposition identity}
\end{align}
which is shown by virtue of the relation (\ref{eq:D-brane_Buscher_rule:4kos}).
Note that $E^{ij}=t^{ij}$.
The identity (\ref{dettDBI}) is an analogue of these, and also used in \cite{Myers:1999ps}.

As a consequence of (\ref{splitting}), we have 
$\det (L^*) = \det (T^* M) \otimes {\det}^{-2} ({\rm Ann}(\Delta))$.
Corresponding to this relation of the determinant bundle, 
we can find relations between the determinants and 
among all there is an important identity
\begin{equation}
\sqrt{\det s} = \sqrt{\det g} \,\det t^{ij},
\label{Giveon identity}
\end{equation}
which we have used in the proof of the invariance of the DBI action in (\ref{detgt2}), 
proved in the appendix \ref{subsec:identity}.
Essentially the same identity appears in the context of T-duality in \cite{Giveon:1994fu}.

The most important decomposition relating to the invariant effective action
is the following combination 
\begin{equation}
{\det}(T^*M) \otimes {\det} (L^*) ={\det}^2 (\Delta^*) \otimes {\det}({\rm Ann}(\Delta)) \otimes {\det}^{-1}({\rm Ann}(\Delta)),
\label{another consequence}
\end{equation}
Using this type of combination of the determinants 
we can construct the object which transforms 
as ${\det} (\Delta^*)$ and consequently has a desired transformation property, 
being invariant under the worldvolume diffeomorphism,
and a diffeomorphism transformation generated by $\epsilon_\perp =\epsilon^i \partial_i$ cancels. 
Therefore, an integrand of the effective action on the worldvolume should belong to this bundle.
Still there are many choices of sections on $\det(T^*M)$ and $\det(L^*)$.
Any combinations of the one from ($\det g,\det E$) and the one from $(\det t,\det t_F,
\det s,
\det(s-as^{-1}a),\det s_F)$ are the candidate.
We listed in the appendix \ref{subsec:app_gene_lie_deri_various_str:3kfs} the transformation of all those determinants.

Among possible tensors, it turns out that $\sqrt{g} \in \det (T^*M)$ and $\sqrt{s_{\cal F}} \in \det (L^*)$, where $s_{\cal F}$ is given in (\ref{def s_F}), are the correct tensors, that is, the last two factors in (\ref{another consequence}) are canceled with each other as we saw in the previous section.
Another good candidate is to take $\det t_{\cal F}$ instead of $\det s_{\cal F}$ from the $\det (L^*)$
but it is not invariant under the full diffeomorphism.

\subsection{Non-linear realized Poincar\'e symmetry}
\label{sec:Poincare_DBI}
Here we fix a generalized Riemannian structure, $g+B$, as a background.
In this case, only a generalized isometry, a subgroup of ${\rm Diff}(M)\ltimes \Omega^2_{\rm closed}(M)$, acts as a symmetry.
A generalized vector $\epsilon +\Lambda  \in{\mathbb T}M$ is called a generalized Killing vector if ${\cal L}_\epsilon  g=0$ and ${\cal L}_\epsilon  B =d\Lambda$ (that is ${\cal L}_{\epsilon +\Lambda} (g+B)=0$) \cite{Grana:2008yw}.
They generate a Lie algebra of the isometry group%
\footnote{Generalized Lie derivatives satisfy in general $[{\cal L}_{\epsilon_1 +\Lambda_1},{\cal L}_{\epsilon_2 +\Lambda_2}]={\cal L}_{[\epsilon_1+\Lambda_1 , \epsilon_2 +\Lambda_2 ]}$.}.
In the case of $dB =0$, the B-field gauge parameter $\Lambda$ is not independent, and is  determined as $\Lambda =i_{\epsilon} B+d\lambda$ up to an arbitrary function $\lambda$.
The Lie algebra relation is then written as
$[{\cal L}_{\epsilon_1 +i_{\epsilon_1} B},{\cal L}_{\epsilon_2 +i_{\epsilon_2} B }]={\cal L}_{[\epsilon_1, \epsilon_2]+i_{[\epsilon_1, \epsilon_2]}B }$.
As already stated, the term $d\lambda$ plays no role in the generalized Lie derivative.

In order to reproduce the result of \cite{Gliozzi:2011hj}, take the Lorentzian signature here and fix $g+B$ to be the Minkowski metric $g=\eta$ and $B=0$, which is a vacuum for string theory.
Then, its isometry group is the Poincar\'e group $ISO(1,D-1)$.
We parametrize an infinitesimal transformation as
\begin{align}
\epsilon  =\epsilon^M\partial_M = (\rho^M + \omega^M{}_N x^N)\partial_M,
\label{Poincare parameter}
\end{align}
where $\rho^M$ is a translation and $\omega^M{}_N$ is a Lorentz transformation,
satisfying $\omega_{MN}+\omega_{NM}=0$.

In the presence of a generalized product structure, ${\rm Diff}(M)$ is broken to ${\rm Fdiff}(M,L)\cap {\rm Fdiff}(M,L^*)$.
Correspondingly, a generalized vector (\ref{Poincare parameter}) satisfying the conditions $\partial_a \epsilon^i=0=\partial_i \epsilon^a$ is an unbroken isometry.
It leads to $\omega_{ai}=0$ and thus $ISO(1,D-1)$ is broken to $ISO(1,p) \times ISO(D-1-p)$ of the form
\begin{align}
\epsilon  =(\rho^a + \omega^a{}_b x^ b)\partial_a+ (\rho^i+\omega^i{}_j x^ j)\partial_i.
\label{unbroken parameter1}
\end{align}
Furthermore, by fixing a leaf at $x^i=0$ as a D-brane, (\ref{Poincare parameter}) should satisfy $\epsilon^i(x^i=0)=0$.
These conditions kill translations $\rho^i$ and we are left with the unbroken group $ISO(1,p) \times SO(D-1-p)$, generated by
\begin{align}
\epsilon  =(\rho^a + \omega^a{}_b x^ b)\partial_a+ \omega^i{}_j x^ j\partial_i.
\label{unbroken paramete2}
\end{align}

As we have seen, a tensor with respect to the Dirac structure $L$ transforms inhomogeneously under of ${\rm Diff}(M)\ltimes \Omega^2_{\rm closed}(M)$.
This is still true for the Poincar\'e subgroup.
This is precisely the non-linearly realized Poincar\'e symmetry.
In fact, by substituting (\ref{Poincare parameter}) in (\ref{eq:gene_Liederi_Aphi:1g6j}), 
we reproduce the result of \cite{Gliozzi:2011hj}.
In particular the broken symmetry given by $\omega_{ai}$ and $\rho^i$ is non-linearly realized.

Finally, we comment on the B-field gauge transformation.
One may also fix only $g$ but take $B$ as a dynamical field.
Note that on a subbundle $C_+$, a B-field gauge transformation is also seen as a shift of $B$, that is,
$B\to B+d\Lambda $.
This is why it is usually called a gauge transformation for $B$.
If, moreover, the $L$-preserving condition $\partial_{[a} \Lambda_{b]} =0$ is satisfied, then the components of $B$ along a leaf is invariant $B_{ab}\to B_{ab}$.
This is usually said that ``$B_{ab}$ cannot be gauged away." 
It is the spontaneous symmetry breaking of the B-field gauge transformation given by $\Lambda_a$.
To restore the broken $\Lambda_a$ transformation, we need to add a gauge field $A_a$, which transforms according to (\ref{eq:gene_Liederi_Aphi:1g6j}).

\section{Conclusion and Discussion}
\label{sec:conclusion}

In this paper, we describe a D-brane in the framework of generalized geometry by choosing the static gauge 
and clarify the structure of the symmetry and its breaking.
By introducing a generalized product structure to formulate the static gauge in this framework, 
scalar fields $\Phi^i$ and a gauge field $A_a$ are unified into a single object $\Phi+A$. 
This object can be regarded either as a diffeomorphism and B-field gauge transformation parameter, 
or as a generalized connection.

We have shown that both scalar fields and gauge field have a shift term in their transformation law, 
from which we conclude that they are NG-bosons for broken translational and B-field gauge transformation, respectively.
We also found that the $U(1)$ gauge symmetry appears as a trivial part of the B-field gauge symmetry.
It is plausible that the same conclusion can be justified in the language of field theory. There is already 
such an analysis of the fluctuations around classical solutions in supergravity theory 
\cite{Adawi:1998ta}.
That a gauge field is simultaneously a NG-boson is also known as the St\"uckelberg mechanism, and in fact, there are 
discussions of spontaneously broken symmetry in the field theory of a $3$-form H-field coupled with a $U(1)$ 
gauge field given in the refs. \cite{Rey:1989ti,Yokoi:2000en,Higashijima:2001sq}.
Since they discuss in the different setting, the relation to our result still has to be clarified.

We have focused on the DBI action in this paper, and shown its invariance under the full spacetime symmetry
and also discussed that these symmetries give some restrictions on the combination of the determinant of possible metrics.
The DBI action is used as a starting point to study BPS condition for a D-brane 
in a generalized calibration in \cite{Koerber:2005qi,Martucci:2005ht,Martucci:2006ij,Koerber:2006hh,Lust:2010by}, 
where the DBI action comes form string theory calculation.
Our discussion here supports this assumption within the generalized geometry.

Note that the non-linearly realized symmetry can determine the functional form 
for $\Phi+A$ and $g+B$ only. 
To determine the overall factor $T_p$ (tension), the dilaton factor $e^{-\phi}$ 
and the factor $2\pi \alpha'$ in front of $F$, other inputs are needed.
In some papers \cite{Grana:2008yw,Jeon:2011cn} a $O(D,D)$ invariant dilaton $d$ is introduced by
\begin{equation}
e^{-2d} = e^{-2\phi}\sqrt{\det g}
\end{equation}
where $\phi$ is an ordinary physical dilaton field which appears in string theory, 
and which transforms under the T-duality as $\phi \to \phi - \frac{1}{2}\ln \det E_{ij}$
\cite{Giveon:1994fu}. 
The new dilaton $e^{-2d}$ also serves as volume element in the target space \cite{Jeon:2011cn}. 
Thus, if we introduce the new dilaton instead of ${\det}^{\frac{1}{4}} g$,
then the correct dilaton term $e^{-\phi}$ can also be recovered.
Because our argument does not take into account T-dualities, we cannot distinguish these two possibilities at this stage.
But in the Ramond-Ramond coupling to D-branes, the dilaton $\phi$ is naturally 
defined as a part of an $O(D,D)$ spinor \cite{Grana:2008yw}.

We expect that our result on the non-linear transformations leads to some constraints 
which can determine possible higher derivative corrections to the DBI action, 
and also can be used to analyze the Chern-Simon term describing the RR-coupling.
The study on the RR-coupling in the context of generalized geometry is initiated by \cite{Grange:2005nm}.
The non-abelian extension of the present formulation, corresponding to the system with multiple D-branes,
 is also an important open problem.  It must be a non-abelian version of the DBI action, however, there is
 no well-established effective action known.

To write the DBI action, we used various identities on determinants.
A similar kind of fractional determinants is found in \cite{Jurco:2012yv,Schupp:2012nq} 
in the context of M-theory branes, where the technique developed 
for the noncommutativity and Seiberg-Witten map was essential.
It is interesting whether our description of D-branes in terms 
of generalized geometry would have some relation to the noncommutative geometry. 

Although this paper is restricted to the local description, its global version 
can also be considered, where the global existence of foliated structures will play an important role.
The classification problem of possible D-branes for a given target space 
in the context of the generalized geometry is an interesting subject as well. 
In the presence of a generalized complex structure, stable branes, called generalized complex branes, 
have been studied by several authors initiated by \cite{Gualtieri:2007bq,Grange:2005nm}.
It should also be compared with the known characterization of multiple D-branes by means of K-homology \cite{Asakawa:2001vm}, which does not rely on the complex structure.

\section*{Acknowledgement}
Authors would like to thank P.~Bouwknegt, V.~Mathai, P.~Schupp, D.~Baraglia, N.~Ikeda, M.~Sato, P.~Kao and F.~Lizzi for fruitful discussions during their stay at Tohoku University.
S.~S. benefited from discussion with  K.~Lee during the Asian Winter school, and 
S.~W. would like to thank B.~Jurco and R.~Szabo for valuable discussions and comments.
We would also like to thank the members of the particle theory and cosmology group, in particular  U.~Carow-Watamura for helpful comments and discussions. 
This work of T.~A. and S.~S. is supported by the GCOE program ``Weaving Science Web beyond Particle-Matter Hierarchy'' at Tohoku University.

\appendix
\section{Appendix}
\label{sec:Some_computations:4kfg}

\subsection{Condition for the Dirac structure}
\label{subsec:Dorfman_involutive}

Here, we derive the condition that the graph $L_{\cal F}=L+{\cal F}(L)$ is a Dirac structure.
As stated in \S \ref{subsec:gene_conn}, this condition was already given in \cite{Liu:1997fj} and 
we reproduce it here using the local coordinate.

Recall that the sections of the subbundle $L_{\cal F}=L+{\cal F}(L)$ have the form
\begin{align}
V &= v^a (x) (\partial_a + \partial_a \Phi^i (x) \partial_i+F_{ab}(x)dx^b) + \xi_i (x) (dx^i - \partial_a \Phi^i (x) dx^i)\nonumber\\
&=e^F (v^a (x) \tilde{\partial}_a  + \xi_i (x) dy^i ),
\label{eq:gene_tangent_vector_Lie:6g5f}
\end{align}
where we allow components $v^a$ and $\xi^i$ as well as $\Phi^i$ and $A_a$ 
to depend on both coordinates $x^a$ and $x^i$.
In the second line, we used the notation introduced in section 2.
Since $L_{\cal F}$ is maximally isotropic (almost Dirac structure), it is sufficient to require the 
Dorfman bracket to be involutive in order to be a Dirac structure.
 
First, we examine the Lie-involutive condition. 
Namely, if $L_{\cal F}$ is involutive, then under the anchor map its image 
$\rho: L_{\cal F}\to TM$ should also be involutive with respect to the Lie bracket of $TM$.
For tangent vectors of the form
\begin{equation}
\rho(V)=v^a(x)(\partial_a + \partial_a \Phi^i (x) \partial_i)=v^a (x)\tilde{\partial}_a,
\label{eq:tangent_vector_Lie:93kd}
\end{equation}
the Lie bracket in $TM$ is given by
\begin{equation}
\begin{split}
[\rho(V) , \rho(V^\prime)]&= [v^a\tilde{\partial}_a, v^{\prime b}\tilde{\partial}_b] \\
&= (v^a \tilde{\partial}_a v^{\prime b} - v^{\prime a} \tilde{\partial}_a v^b)\tilde{\partial}_b
+ v^a v^{\prime b} [\tilde{\partial}_a,\tilde{\partial}_b].
\end{split}
\end{equation}
Since $[\tilde{\partial}_a,\tilde{\partial}_b]=(\partial_a \Phi^j\partial_j \partial_b \Phi^i -
\partial_b \Phi^j\partial_j \partial_a \Phi^i)\partial_i $ is not proportional to $\tilde{\partial}_a$, 
the last term should vanish in order for $\rho(L)$ to be closed under the Lie bracket.
Therefore, the condition is
\begin{equation}
\partial_i (\partial_b \Phi^j) = 0 ,
\label{Lie involutive condition}
\end{equation}
i.e. the field $\partial_a\Phi^i$ is independent of $x^i$.

Keeping this in mind, we examine the Dorfman-involutive condition. 
The Dorfman bracket between generalized tangent vectors of the form (\ref{eq:gene_tangent_vector_Lie:6g5f}) is given by
\begin{equation}
\begin{split}
[V, V^\prime] &=[e^F (v^a \tilde{\partial}_a  + \xi_i dy^i), e^F (v^{\prime b} \tilde{\partial}_b  + \xi^\prime_j dy^j)]\\
&= e^F [v^a \tilde{\partial}_a  + \xi_i dy^i, v^{\prime b} \tilde{\partial}_b  + \xi^\prime_j dy^j]-\iota_{v^a \tilde{\partial}_a}\iota_{v^{\prime b} \tilde{\partial}_b} dF\\
&= e^F [v^a \tilde{\partial}_a, v^{\prime b} \tilde{\partial}_b]
+{\cal L}_{v^a \tilde{\partial}_a} \xi^\prime_j dy^j
-\iota_{v^{\prime b} \tilde{\partial}_b} d(\xi_i dy^i)
-\iota_{v^a \tilde{\partial}_a}\iota_{v^{\prime b} \tilde{\partial}_b} dF,
\end{split} \label{eq:Dorfman-involutivity:4qf9}
\end{equation}
where we used the Courant automorphism for a B-transformation given by $e^F$.
The first term lies in $L_{\cal F}$, since $[\tilde{\partial}_a,\tilde{\partial}_b]=0$ as is shown above.
Since $\iota_{v^a \tilde{\partial}_a}dy^i=0$ and $d(dy^i)=0$, 
the second and the third term are written as
\begin{equation}
\begin{split}
{\cal L}_{v^a \tilde{\partial}_a} \xi^\prime_j dy^j
-\iota_{v^{\prime b} \tilde{\partial}_b} d(\xi_i dy^i)
&=\iota_{v^a \tilde{\partial}_a} d(\xi^\prime_j dy^j)
-\iota_{v^{\prime b} \tilde{\partial}_b} d(\xi_i dy^i)\\
&=v^{a} \tilde{\partial}_a \xi^\prime_j dy^j -v^{\prime b} \tilde{\partial}_b\xi_i dy^i,
\end{split}
\end{equation}
which is also an element in $L_{\cal F}$.
By using the fact that the exterior derivative of an arbitrary form $\omega$
is 
\begin{align}
d\omega= dx^a\partial_a \omega+ dx^i \partial_i \omega
=dx^a\tilde{\partial}_a \omega+ dy^i \partial_i \omega~,
\end{align}
the last term in (\ref{eq:Dorfman-involutivity:4qf9}) can be written as
\begin{equation}
\begin{split}
-\iota_{v^a \tilde{\partial}_a}\iota_{v^{\prime b} \tilde{\partial}_b} dF
&=-\iota_{v^a \tilde{\partial}_a}\iota_{v^{\prime b} \tilde{\partial}_b}
\tilde{\partial}_c F_{de} dx^c\wedge dx^d \wedge dx^e -v^a v^{\prime b} \partial_i F_{ab}dy^i.
\end{split}
\end{equation}
Although the second term lies in $L_{\cal F}$, the first term does not.
Requiring the first term to vanish for arbitrary $v^a$ and $v^{\prime b}$, 
we obtain the following condition:
\begin{align}
{\tilde \partial}_a F_{bc} + {\tilde \partial}_b F_{ca} + {\tilde \partial}_c F_{ab} &= 0. \label{eq:Bianchi appendix}
\end{align}

The two equations (\ref{Lie involutive condition}) and (\ref{eq:Bianchi appendix}) are the conditions 
to guarantee that $L_{\cal F}$ is a Dirac structure.

\subsection{Derivation of the non-linear transformation law}
\label{subsec:derivation}

Let $T \in L^* \otimes L^*$ be a tensor and define its graph by $L_T=L+T(L)$.
Consider an action of a generalized Lie derivative ${\mathcal L}_{\epsilon+\Lambda} (V+T(V))$ 
on a section $V+T(V) \in L_T$ with $V\in L$.
The first term, ${\mathcal L}_{\epsilon+\Lambda} V$ is already evaluated in (\ref{LV}).
For the tensor $T$ with the form
\begin{equation}
T = T_{ab}(x) dx^a \otimes dx^b + T_a^{\ j}(x) dx^a \otimes \partial_j + T^i_{\ b}(xi) \partial_i \otimes dx^b + T^{ij}(x) \partial_i \otimes \partial_j ,
\end{equation}
the generalized Lie derivative acts on a tensor, like in the ordinary case, as
\begin{equation}
\begin{split}
{\mathcal L}_{\epsilon+\Lambda}(T_{ab}dx^a\otimes dx^b) 
&= \epsilon^M\partial_M T_{ab} dx^a \otimes dx^b+\partial_M \epsilon^c T_{cb} dx^M \otimes dx^b \\
&\quad + T_{ac}\partial_N \epsilon^c dx^a \otimes dx^N, \\
{\mathcal L}_{\epsilon+\Lambda}(T_a^{\ j} dx^a \otimes \partial_j) 
&= \epsilon^M\partial_M T_a^{\ j} dx^a \otimes \partial_j + \partial_M \epsilon^c T_c^{\ j} dx^M \otimes \partial_j \\
& \quad -T_a^{\ k}\partial_k\epsilon^M dx^a\otimes\partial_M -T_a^{\ k} \partial_{[k} \Lambda_{N]} dx^a\otimes dx^N, \\
{\mathcal L}_{\epsilon+\Lambda}(T^i_{\ b} \partial_i \otimes dx^b) 
&=\epsilon^M\partial_M T^i_{\ b} \partial_i\otimes dx^b- \partial_k\epsilon^M T^k_{\ b} \partial_M\otimes dx^b \\
&\quad + T^i_{\ c} \partial_N\epsilon^c \partial_i \otimes dx^N - \partial_{[k} \Lambda_{M]} T^k_{\ b} dx^M\otimes dx^b, \\
{\mathcal L}_{\epsilon+\Lambda}(T^{ij}\partial_i \otimes \partial_j)&= \epsilon^M\partial_M T^{ij}\partial_i\otimes \partial_j-\partial_k\epsilon^M T^{kj}\partial_M\otimes\partial_j-T^{ik}\partial_k\epsilon^N \partial_i\otimes\partial_N \\
&\quad-\partial_{[k}\Lambda_{M]} T^{kj}dx^M\otimes\partial_j -T^{ik}\partial_{[k}\Lambda_{N]}\partial_i\otimes dx^N.
\end{split}
\end{equation}
By using these relations, ${\mathcal L}_{\epsilon+\Lambda} (V+T(V))$ 
can be written in the form of (\ref{eq:def_liederoffield:3ld0}), or more explicitly, 
by splitting $\delta V$ into the vector and $1$-form as $\delta V=\delta v+\delta \xi$
\begin{equation}
\begin{split}
{\mathcal L}_{\epsilon+\Lambda} (V+T(V)) &= (-\delta v^a)(\partial_a+T_a^{\ j}\partial_j+T_{ab}dx^b) +(-\delta\xi_i)(dx^i+T^i_{\ b}dx^b+T^{ij}\partial_j) \\
&\quad +v^a\{(-\delta T_a^{\ j})\partial_j+(-\delta T_{ab})dx^b\}+\xi_i\{(-\delta T^i_{\ b})dx^b+(-\delta T^{ij})\partial_j\}.
\label{form of the main text}
\end{split}
\end{equation}
Here the first line is the part of the graph of the $\delta V$ in the form of $\delta V + T(\delta V)$, with
\begin{equation}
\begin{split}
\delta v^a &= -\epsilon^M\partial_Mv^a +v^b\partial_b\epsilon^a + v^b T_b^{\ k} \partial_k\epsilon^a+\xi_jT^{jk}\partial_k\epsilon^a, \\
\delta \xi_i &= -\epsilon^M\partial_M\xi_i-\xi_k\partial_i\epsilon^k+v^b\partial_{[b}\Lambda_{i]} -v^bT_{bc}\partial_i\epsilon^c + v^bT_b^{\ k}\partial_{[k}\Lambda_{i]} \\
& \quad -\xi_j T^j_c \partial_i \epsilon^c + \xi_j T^{jk} \partial_{[k}\Lambda_{i]}.
\end{split}
\end{equation}
The second line in (\ref{form of the main text}) defines 
the non-linear transformation law for the tensor $T$, where
\begin{subequations}
\begin{align}
\delta T_{ab} &= -\epsilon^M \partial_M T_{ab} - T_{ac}\partial_b \epsilon^c -\partial_a \epsilon^c T_{cb} +\partial_{[k} \Lambda_{a]} T^k_{\ b} +T_a^{\ k}\partial_{[k} \Lambda_{b]} \notag\\
& \quad -T_a^{\ k}\partial_k \epsilon^c T_{cb}+T_{ac}\partial_k\epsilon^c T^k_{\ b} -T_a^{\ k}\partial_{[k}\Lambda_{l]} T^{l}_{\ b} +\partial_{[a}\Lambda_{b]}, \\
\delta T_a^{\ j} &= -\epsilon^M\partial_M T_a^{\ j} -\partial_a \epsilon^c T_c^{\ j} +T_a^{\ k}\partial_k \epsilon^j  +\partial_{[k}\Lambda_{a]} T^{kj} \notag \\
& \quad -T_a^{\ k}\partial_k \epsilon^c T_{c}^{\ j} +T_{ac}\partial_k\epsilon^c T^{kj} -T_a^{\ k}\partial_{[k}\Lambda_{l]} T^{lj} +\partial_a \epsilon^j, \\
\delta T^i_{\ b} &= -\epsilon^M\partial_M T^i_{\ b} +\partial_k \epsilon^i T^k_{\ b} -T^i_{\ c}\partial_b \epsilon^c  + T^{ik}\partial_{[k}\Lambda_{b]} \notag \\
& \quad -T^{ik}\partial_k \epsilon^c T_{cb} +T^i_{\ c}\partial_k\epsilon^c T^k_{\ b} -T^{ik}\partial_{[k}\Lambda_{l]}T^{l}_{\ b} -\partial_b \epsilon^i, \\
\delta T^{ij} &=-\epsilon^M \partial_M T^{ij} +\partial_k \epsilon^i T^{kj} +T^{ik}\partial_k \epsilon^j \notag\\
& \quad -T^{ik} \partial_k \epsilon^c T_c^{\ j} +T^i_{\ c} \partial_k \epsilon^c T^{kj} -T^{ik} \partial_{[k}\Lambda_{l]}T^{lj}.
\end{align} \label{eq:Lie_field_LL:4e3da}
\end{subequations}

\subsection{Generalized metric as various graphs}
\label{subsec:app_genemet_as_graphs}

Here we derive the formula (\ref{eq:D-brane_Buscher_rule:4kos}), which relates two different tensors describing a same generalized Riemannian structure. 
A generalized vector field $V_+$ in the 
positive-definite subbundle $C_+$ is written either by using $v^M \partial_M \in TM$,
\begin{equation}
V_+ = v^M (\partial_M + E_{MN}dx^N) \quad \in C_+,
\label{eq_app:Gene_metricasgraph:4kfo}
\end{equation}
as a graph of $E=g+B : TM \to T^\ast M$, or is written by using 
$w^a \partial_a +\xi_i dx^i \in L$,
\begin{equation}
V_+ = w^a(\partial_a + t_a^{\ j}\partial_j + t_{ab}dx^b) + \xi_i (dx^i+t^i_{\ b}dx^b +t^{ij}\partial_j) \quad \in C_+,
\label{eq_app:GMgraL0_v9ff}
\end{equation}
as a graph of $t: L \to L^\ast$. 
By comparing (\ref{eq_app:Gene_metricasgraph:4kfo}) and (\ref{eq_app:GMgraL0_v9ff}), 
we obtain the following relations 
\begin{align}
w^a&=v^a, \label{eq_app:GMcompTXL1_aic5} \\
w^at_a^{\ j}+\xi_i t^{ij} &= v^j, \label{eq_app:GMcompTXL2_moe}\\
w^at_{ab}+\xi_i t^i_{\ b} &= v^M E_{Mb}, \label{eq_app:GMcompTXL3_4bv}\\
\xi_j &= v^M E_{Mj}, \label{eq_app:GMcompTXL4_sxr5}
\end{align}
where we have used the independence of the basis $\{\partial_a, \partial_i, dx^a, dx^i\}$.
Substituting eqs. (\ref{eq_app:GMcompTXL1_aic5}) and (\ref{eq_app:GMcompTXL4_sxr5}) 
into eqs.(\ref{eq_app:GMcompTXL2_moe}) and (\ref{eq_app:GMcompTXL3_4bv}), the above relations lead to
\begin{equation}
\begin{split}
v^a (t_a^{\ j} +E_{ak}t^{kj})+v^i(E_{ik}t^{kj}-\delta_i^j) &= 0, \\
v^a (t_{ab}+E_{ak}t^k_{\ b}-E_{ab})+v^i(E_{ik} t^k_{\ b} -E_{ib}) &= 0.
\end{split}
\end{equation}
Since these equations must hold for arbitrary $v^a$ and $v^i$, we obtain the condition 
for the tensor $t\in L^* \otimes L^*$ as :
\begin{equation}
\begin{split}
t^{ij} &= E^{ij}, \qquad \qquad t_a^{\ j}= - E_{ak}E^{kj}, \\
t^i_{\ b} &= E^{ik}E_{kb}, \qquad \  t_{ab} = E_{ab} - E_{ak} E^{kl} E_{lb},
\end{split}
\end{equation}
where $E^{ij}$ is the inverse of $E_{ij}$ satisfying $E^{ij}E_{jk}=\delta^i_k$,
existence of which is guaranteed by the positive definiteness of $C_+$.
These are the equations given in (\ref{eq:D-brane_Buscher_rule:4kos}).

\subsection{Consistency between linear and non-linear transformation laws}
\label{subsec:consistency}

We show here the consistency,i.e.,the commutativity of the diagram (\ref{commutativeDG})
 stated in \S \ref{sec:genelie_metric:e3lf}.
To this end, we take the ``right-down"  route $E \to E+\delta E \to t+\delta t$, and then compare with 
(\ref{eq:Lie_of_genemet_tensor:ek4l}), which coincides with the ``down-right" route.

The action of a generalized Lie derivative $-{\cal L}_{\epsilon +\Lambda}$ on the tensor $E=g+B$ is 
\begin{equation}
\delta E_{MN} = - \epsilon^L \partial_L E_{MN} -\partial_M \epsilon^L E_{LN} - E_{ML}\partial_N\epsilon^L + \partial_{[M}\Lambda_{N]}.
\end{equation}
This defines $E+\delta E$.
We also need the action on the inverse $E^{ij}$:
\begin{equation}
\delta E^{ij} = -E^{ik}\delta E_{kl}E^{lj} =E^{ik}(\epsilon^L \partial_L E_{kl} + \partial_k \epsilon^L E_{Ll} + E_{kL}\partial_l\epsilon^L - \partial_{[k}\Lambda_{l]})E^{lj},
\end{equation}
which is derived from $\delta(E_{ik}E^{kj})=\delta E_{ik}E^{kj} + E_{ik}\delta E^{kj}=0$.

Substituting $E+\delta E$ to the relations (\ref{eq:D-brane_Buscher_rule:4kos}), we obtain 
$t +\delta t$, where
\begin{align}
\begin{split}
\delta t_{ab} &= \delta E_{ab} - \delta E_{ak}E^{kl}E_{lb} -E_{ak}\delta E^{kl}E_{lb} - E_{ak}E^{kl}\delta E_{lb} \\
&= - \epsilon^L \partial_L E_{ab} -\partial_a \epsilon^L E_{Lb} - E_{aL}\partial_b\epsilon^L + \partial_{[a}\Lambda_{b]} \\
&\quad -( - \epsilon^L \partial_L E_{ak} -\partial_a \epsilon^L E_{Lk} - E_{aL}\partial_k\epsilon^L + \partial_{[a}\Lambda_{k]}) E^{kl}E_{lb} \\
&\quad -E_{ai}E^{ik}(\epsilon^L \partial_L E_{kl} + \partial_k \epsilon^L E_{Ll} + E_{kL}\partial_l\epsilon^L - \partial_{[k}\Lambda_{l]})E^{lj}E_{jb} \\
& \quad -E_{ak}E^{kl}(- \epsilon^L \partial_L E_{lb} -\partial_l \epsilon^L E_{Lb} - E_{lL}\partial_b\epsilon^L + \partial_{[l}\Lambda_{b]}) \\
&= -\epsilon^M \partial_M t_{ab} - t_{ac}\partial_b \epsilon^c -\partial_a \epsilon^c t_{cb} +\partial_{[k} \Lambda_{a]} t^k_{\ b} +t_a^{\ k}\partial_{[k} \Lambda_{b]} \notag\\
& \quad -t_a^{\ k}\partial_k \epsilon^c t_{cb}+t_{ac}\partial_k\epsilon^c t^k_{\ b} -t_a^{\ k}\partial_{[k}\Lambda_{l]} t^{l}_{\ b} +\partial_{[a}\Lambda_{b]},
\end{split}\\
\begin{split}
\delta t_a^{\ j} &= - \delta E_{ak}E^{kj} - E_{ak}\delta E^{kj} \\
&= -(- \epsilon^L \partial_L E_{ak} -\partial_a \epsilon^L E_{Lk} - E_{aL}\partial_k\epsilon^L + \partial_{[a}\Lambda_{k]})E^{kj} \\
&\quad - E_{ai}E^{ik}(\epsilon^L \partial_L E_{kl} + \partial_k \epsilon^L E_{Ll} + E_{kL}\partial_l\epsilon^L - \partial_{[k}\Lambda_{l]})E^{lj} \\
&= -\epsilon^M\partial_M t_a^{\ j} -\partial_a \epsilon^c t_c^{\ j} +t_a^{\ k}\partial_k \epsilon^j  +\partial_{[k}\Lambda_{a]} t^{kj} \notag \\
& \quad -t_a^{\ k}\partial_k \epsilon^c t_{c}^{\ j} +t_{ac}\partial_k\epsilon^c t^{kj} -t_a^{\ k}\partial_{[k}\Lambda_{l]} t^{lj} +\partial_a \epsilon^j,
\end{split}\\
\begin{split}
\delta t^i_{\ b} &= \delta E^{ik}E_{kb} + E^{ik} \delta E_{kb} \\
&= E^{ik}(\epsilon^L \partial_L E_{kl} + \partial_k \epsilon^L E_{Ll} + E_{kL}\partial_l\epsilon^L - \partial_{[k}\Lambda_{l]})E^{lj} E_{jb} \\
&\quad + E^{ik}(- \epsilon^L \partial_L E_{kb} -\partial_k \epsilon^L E_{Lb} - E_{kL}\partial_b\epsilon^L + \partial_{[k}\Lambda_{b]})\\
&= -\epsilon^M\partial_M t^i_{\ b} +\partial_k \epsilon^i t^k_{\ b} -t^i_{\ c}\partial_b \epsilon^c  + t^{ik}\partial_{[k}\Lambda_{b]} \notag \\
& \quad -t^{ik}\partial_k \epsilon^c t_{cb} +t^i_{\ c}\partial_k\epsilon^c t^k_{\ b} -t^{ik}\partial_{[k}\Lambda_{l]}t^{l}_{\ b} -\partial_b \epsilon^i,
\end{split}\\
\begin{split}
\delta t^{ij} &= \delta E^{ij} = E^{ik}(\epsilon^L \partial_L E_{kl} + \partial_k \epsilon^L E_{Ll} + E_{kL}\partial_l\epsilon^L - \partial_{[k}\Lambda_{l]})E^{lj} \\
&= -\epsilon^M \partial_M t^{ij} +\partial_k \epsilon^i t^{kj} +t^{ik}\partial_k \epsilon^j \notag\\
& \quad -t^{ik} \partial_k \partial^c t_c^{\ j} +t^i_{\ c} \partial_k \epsilon^c t^{kj} -t^{ik} \partial_{[k}\Lambda_{l]}t^{lj}.
\end{split}
\end{align}
These transformation rules agree with (\ref{eq:Lie_of_genemet_tensor:ek4l}).

\subsection{Non-linear transformation laws for various determinants}
\label{subsec:app_gene_lie_deri_various_str:3kfs}

We here summarize non-linear transformation laws for various determinants, used in this paper.
For this end, we need the transformation law for the symmetric/anti-symmetric part of the tensor $t=s+a$, given by
\begin{subequations}
\begin{align}
\delta s_{ab} &= -\epsilon^M \partial_M s_{ab} - s_{ac}\partial_b \epsilon^c -\partial_a \epsilon^c s_{cb} +\partial_{[k} \Lambda_{a]} s^k_{\ b} +s_a^{\ k}\partial_{[k} \Lambda_{b]} \notag\\
& \quad -s_a^{\ k}\partial_k \epsilon^c a_{cb}-a_a^{\ k}\partial_k \epsilon^c s_{cb}+s_{ac}\partial_k\epsilon^c a^k_{\ b}+a_{ac}\partial_k\epsilon^c s^k_{\ b} \\ 
& \quad -s_a^{\ k}\partial_{[k}\Lambda_{l]} a^{l}_{\ b}-a_a^{\ k}\partial_{[k}\Lambda_{l]} s^{l}_{\ b}, \notag \\
\delta s_a^{\ j} &= -\epsilon^M\partial_M s_a^{\ j} -\partial_a \epsilon^c s_c^{\ j} +s_a^{\ k}\partial_k \epsilon^j  +\partial_{[k}\Lambda_{a]} s^{kj} \notag \\
& \quad -s_a^{\ k}\partial_k \epsilon^c a_{c}^{\ j}-a_a^{\ k}\partial_k \epsilon^c s_{c}^{\ j} +s_{ac}\partial_k\epsilon^c a^{kj} +a_{ac}\partial_k\epsilon^c s^{kj} \\
& \quad -s_a^{\ k}\partial_{[k}\Lambda_{l]} a^{lj}-a_a^{\ k}\partial_{[k}\Lambda_{l]} s^{lj}, \notag \\
\delta s^i_{\ b} &= -\epsilon^M\partial_M s^i_{\ b} +\partial_k \epsilon^i s^k_{\ b} -s^i_{\ c}\partial_b \epsilon^c  + s^{ik}\partial_{[k}\Lambda_{b]} \notag \\
& \quad -s^{ik}\partial_k \epsilon^c a_{cb} -a^{ik}\partial_k \epsilon^c s_{cb} +s^i_{\ c}\partial_k\epsilon^c a^k_{\ b} +a^i_{\ c}\partial_k\epsilon^c s^k_{\ b} \\
& \quad -s^{ik}\partial_{[k}\Lambda_{l]}a^{l}_{\ b} -a^{ik}\partial_{[k}\Lambda_{l]}s^{l}_{\ b}, \notag \\
\delta s^{ij} &=-\epsilon^M \partial_M s^{ij} +\partial_k \epsilon^i s^{kj} +s^{ik}\partial_k \epsilon^j \notag\\
& \quad -s^{ik} \partial_k \epsilon^c a_c^{\ j} -a^{ik} \partial_k \epsilon^c s_c^{\ j} +s^i_{\ c} \partial_k \epsilon^c a^{kj} +a^i_{\ c} \partial_k \epsilon^c s^{kj} \\
&\quad -s^{ik} \partial_{[k}\Lambda_{l]}a^{lj} -a^{ik} \partial_{[k}\Lambda_{l]}s^{lj}, \notag
\end{align} \label{eq:geneLie_of_s:7kda}
\end{subequations}
and 
\begin{subequations}
\begin{align}
\delta a_{ab} &= -\epsilon^M \partial_M a_{ab} - a_{ac}\partial_b \epsilon^c -\partial_a \epsilon^c a_{cb} +\partial_{[k} \Lambda_{a]} a^k_{\ b} +a_a^{\ k}\partial_{[k} \Lambda_{b]} \notag\\
& \quad -s_a^{\ k}\partial_k \epsilon^c s_{cb} -a_a^{\ k}\partial_k \epsilon^c a_{cb} +s_{ac}\partial_k\epsilon^c s^k_{\ b} +a_{ac}\partial_k\epsilon^c a^k_{\ b} \\ 
& \quad -s_a^{\ k}\partial_{[k}\Lambda_{l]} s^{l}_{\ b} -a_a^{\ k}\partial_{[k}\Lambda_{l]} a^{l}_{\ b} +\partial_{[a}\Lambda_{b]}, \notag \\
\delta a_a^{\ j} &= -\epsilon^M\partial_M a_a^{\ j} -\partial_a \epsilon^c a_c^{\ j} +a_a^{\ k}\partial_k \epsilon^j  +\partial_{[k}\Lambda_{a]} a^{kj} \notag \\
& \quad -s_a^{\ k}\partial_k \epsilon^c s_{c}^{\ j} -a_a^{\ k}\partial_k \epsilon^c a_{c}^{\ j} +s_{ac}\partial_k\epsilon^c s^{kj} +a_{ac}\partial_k\epsilon^c a^{kj} \\
& \quad -s_a^{\ k}\partial_{[k}\Lambda_{l]} s^{lj} -a_a^{\ k}\partial_{[k}\Lambda_{l]} a^{lj} +\partial_a \epsilon^j, \notag\\
\delta a^i_{\ b} &= -\epsilon^M\partial_M a^i_{\ b} +\partial_k \epsilon^i a^k_{\ b} -a^i_{\ c}\partial_b \epsilon^c  + a^{ik}\partial_{[k}\Lambda_{b]} \notag \\
& \quad -s^{ik}\partial_k \epsilon^c s_{cb} -a^{ik}\partial_k \epsilon^c a_{cb} +s^i_{\ c}\partial_k\epsilon^c s^k_{\ b} +a^i_{\ c}\partial_k\epsilon^c a^k_{\ b} \\
& \quad -s^{ik}\partial_{[k}\Lambda_{l]}s^{l}_{\ b} -a^{ik}\partial_{[k}\Lambda_{l]}a^{l}_{\ b} -\partial_b \epsilon^i,\notag\\
\delta a^{ij} &=-\epsilon^M \partial_M a^{ij} +\partial_k \epsilon^i a^{kj} +a^{ik}\partial_k \epsilon^j \notag\\
& \quad -s^{ik} \partial_k \epsilon^c s_c^{\ j} -a^{ik} \partial_k \epsilon^c a_c^{\ j} +s^i_{\ c} \partial_k \epsilon^c s^{kj} +a^i_{\ c} \partial_k \epsilon^c a^{kj} \\
& \quad -s^{ik} \partial_{[k}\Lambda_{l]}s^{lj} -a^{ik} \partial_{[k}\Lambda_{l]}a^{lj}. \notag
\end{align} \label{eq:geneLie_of_a:3s2lg}
\end{subequations}
The non-linear transformation laws for various determinants are summarized as
\begin{subequations}
\begin{align}
\delta \det g &= -\epsilon^M \partial_M \det g +\det g\left\{ -2\partial_M\epsilon^M \right\}, \\
\delta \det s &= -\epsilon^M \partial_M \det s +\det s\left\{ -2\partial_c\epsilon^c +2\partial_k\epsilon^k -4a_c^{\ k}\partial_k\epsilon^c -4a^{kl}\partial_{l} \Lambda_{k} \right\}, \\
\delta \det s_{\mathcal F} &= -\epsilon^M \partial_M \det s_{\mathcal F} +\det s_{\mathcal F} \left\{ -2\partial_c\epsilon^c +2\partial_k\epsilon^k -4{\mathcal F}_c^{\ k}\partial_k\epsilon^c \right\}, \\
\delta \det t &= -\epsilon^M \partial_M \det t + \det t \left[ -2\partial_c\epsilon^c +2\partial_k\epsilon^k -2a_c^{\ k}\partial_k\epsilon^c  -2a^{kl} \partial_{l} \Lambda_{k} \right] \notag \\
&\quad +\det t \left[ (t^{-1})^{ba}\partial_{[a} \Lambda_{b]} +(t^{-1})_j^{\ a} \partial_a\epsilon^j -(t^{-1})^b_{\ i}\partial_b\epsilon^i \right], \\
\delta \det t^{ij} &= -\epsilon^M \partial_M \det t^{ij} + \det t_{ij} \left[ 2\partial_k\epsilon^k -2a_c^{\ k}\partial_k\epsilon^c  -2a^{kl} \partial_{l} \Lambda_{k} \right], \\
\delta \det {t_{\mathcal F}} 
&= -\epsilon^M \partial_M \det {t_{\mathcal F}} + \det {t_{\mathcal F}} \left[ -2\partial_c\epsilon^c +2\partial_k\epsilon^k -2(a_c^{\ k}+{\mathcal F}_c^{\ k})\partial_k\epsilon^c  -2a^{kl} \partial_{l} \Lambda_{k} \right].
\end{align}
\end{subequations}

\subsection{Identities on determinants}
\label{subsec:identity}

Here we prove $3$ identities on determinants given in (\ref{detgt2}), (\ref{dettF2}) and (\ref{dettDBI}) in \S \ref{sec:inv_DBI} and also used in \S \ref{sec:inv_Effective}.
First we show the relation (\ref{detgt2}). 
We rewrite (\ref{eq:D-brane_Buscher_rule:4kos}) 
as a matrix and decompose it into two matrices as
\begin{equation}
\begin{split}
t &= \begin{pmatrix} E_{ab} - E_{ak} E^{kl} E_{lb} & -E_{ak} E^{kj} \\ E^{ik} E_{kb} & E^{ij} \end{pmatrix} \\
&= \begin{pmatrix} \delta_{ac} & E_{ak} \\ 0 & E_{ik} \end{pmatrix}^{-1} \begin{pmatrix} E_{cb} & 0 \\ E_{kb} & \delta_{kj} \end{pmatrix} \\
&\overset{\rm def.}{=} m^{-1} \cdot n,
\end{split}
\end{equation}
where we defined two $D\times D$ matrices $n$ and $m$. 
Then the matrix $s$, the symmetric part of $t$, can also be written as
\begin{equation}
\begin{split}
s = \frac{1}{2} \left(t+ t^T \right) &= \frac{1}{2} \left(m^{-1}n+n^{T} m^{-1T}\right) \\
&= \frac{1}{2} m^{-1} \cdot \left( n m^T+ m n^T \right) \cdot m^{-1T} \\
&= m^{-1} \cdot g \cdot m^{-1T}
\end{split} \label{eq:symmetric_tensor_s:e4g}
\end{equation}
By taking the determinant of (\ref{eq:symmetric_tensor_s:e4g}), we have
\begin{equation}
\det s = (\det m)^{-2} \det g = (\det t^{ij})^2 \det g.
\end{equation}

Next, we show (\ref{dettF2}).
Let us decompose a matrix $t=s+a$ into symmetric/anti-symmetric parts.
If the existence of $s^{-1}$ is assumed, one can show
\begin{align}
\det t &=\left(\det (s+a)\,\det (s+a) \right)^{\frac{1}{2}}\nonumber\\
&=\left(\det s(1+s^{-1}a)\,\det (1+as^{-1})s \right)^{\frac{1}{2}}\nonumber\\
&=\det s\, \left(\det (1+s^{-1}a)(1-s^{-1}a) \right)^{\frac{1}{2}}\nonumber\\
&=\det s\, \left(\det (1-s^{-1}as^{-1}a) \right)^{\frac{1}{2}}\nonumber\\
&=\det s\, \left(\det s^{-1}(s-as^{-1}a) \right)^{\frac{1}{2}}\nonumber\\
&={\det}^{\frac{1}{2}} s\, {\det}^{\frac{1}{2}} (s-as^{-1}a).
\end{align}
Applying this relation to $t_{\cal F}$, we obtain (\ref{dettF2}).

Finally, by using the formula (\ref{submatrix identity}), we find the relation (\ref{dettDBI}) as follows:
\begin{equation}
\begin{split}
\det t_{\mathcal F} &= \det \begin{pmatrix} E_{ab} - E_{ak} E^{kl} E_{lb} - F_{ab} & -E_{ak} E^{kj} - \partial_a \Phi^j \\ E^{ik} E_{kb} + \partial_b \Phi^i & E^{ij} \end{pmatrix} \\
&= \det \left( E_{ab} + \partial_a \Phi^k E_{kb} + E_{ak}\partial_b \Phi^k +\partial_a \Phi^i E_{ij} \partial_b\Phi^j -F_{ab} \right) \det E^{ij} \\
&= \det t^{ij}\det(\varphi_\Phi^\ast (g+B) -F )_{ab} ~.
\end{split}
\end{equation}


\printindex


\begin{thebibliography}{999}

\bibitem{Fradkin:1985qd} 
  E.~S.~Fradkin and A.~A.~Tseytlin,
  ``Nonlinear Electrodynamics from Quantized Strings,''
  Phys.\ Lett.\ B {\bf 163}, 123 (1985).

\bibitem{Abouelsaood:1986gd} 
  A.~Abouelsaood, C.~G.~Callan, Jr., C.~R.~Nappi and S.~A.~Yost,
  ``Open Strings in Background Gauge Fields,''
  Nucl.\ Phys.\ B {\bf 280}, 599 (1987).

\bibitem{Callan:1988wz} 
  C.~G.~Callan, Jr., C.~Lovelace, C.~R.~Nappi and S.~A.~Yost,
  ``Loop Corrections to Superstring Equations of Motion,''
  Nucl.\ Phys.\ B {\bf 308}, 221 (1988).

\bibitem{Tseytlin:1997csa} 
  A.~A.~Tseytlin,
  ``On nonAbelian generalization of Born-Infeld action in string theory,''
  Nucl.\ Phys.\ B {\bf 501}, 41 (1997)
  [hep-th/9701125].

\bibitem{Tseytlin:1999dj} 
  A.~A.~Tseytlin,
  ``Born-Infeld action, supersymmetry and string theory,''
  In *Shifman, M.A. (ed.): The many faces of the superworld* 417-452
  [hep-th/9908105].

\bibitem{Nambu:1960xd} 
  Y.~Nambu,
  ``Axial vector current conservation in weak interactions,''
  Phys.\ Rev.\ Lett.\  {\bf 4}, 380 (1960).

\bibitem{Goldstone:1961eq} 
  J.~Goldstone,
  ``Field Theories with Superconductor Solutions,''
  Nuovo Cim.\  {\bf 19}, 154 (1961).

\bibitem{Low:2001bw} 
  I.~Low and A.~V.~Manohar,
  ``Spontaneously broken space-time symmetries and Goldstone's theorem,''
  Phys.\ Rev.\ Lett.\  {\bf 88}, 101602 (2002)
  [hep-th/0110285].

\bibitem{Aharony:2011gb} 
  O.~Aharony and M.~Dodelson,
  ``Effective String Theory and Nonlinear Lorentz Invariance,''
  JHEP {\bf 1202}, 008 (2012)
  [arXiv:1111.5758 [hep-th]].

\bibitem{Dubovsky:2012sh} 
  S.~Dubovsky, R.~Flauger and V.~Gorbenko,
  ``Effective String Theory Revisited,''
  [arXiv:1203.1054 [hep-th]].

\bibitem{Gomis:2012ki} 
  J.~Gomis, K.~Kamimura and J.~M.~Pons,
  ``Non-linear Realizations, Goldstone bosons of broken Lorentz rotations and effective actions for p-branes,''
  [arXiv:1205.1385 [hep-th]].

\bibitem{Gliozzi:2011hj}
  F.~Gliozzi,
  ``Dirac-Born-Infeld action from spontaneous breakdown of Lorentz symmetry in brane-world scenarios,''
  Phys.\ Rev.\ D {\bf 84}, 027702 (2011)
  [arXiv:1103.5377 [hep-th]].

\bibitem{Casalbuoni:2011fq} 
  R.~Casalbuoni, J.~Gomis and K.~Kamimura,
  ``Space-time transformations of the Born-Infeld gauge field of a D-brane,''
  Phys.\ Rev.\ D {\bf 84}, 027901 (2011)
  [arXiv:1104.4916 [hep-th]].

\bibitem{Hitchin:2004ut} 
  N.~Hitchin,
  ``Generalized Calabi-Yau manifolds,''
  Quart.\ J.\ Math.\ Oxford Ser.\  {\bf 54}, 281 (2003)
  [math/0209099 [math-dg]].

\bibitem{Gualtieri:2003dx}
  M.~Gualtieri,
  ``Generalized complex geometry,''
  [arXiv:math/0401221 [math.DG]].

\bibitem{Zabzine:2004dp} 
  M.~Zabzine,
  ``Geometry of D-branes for general N=(2,2) sigma models,''
  Lett.\ Math.\ Phys.\  {\bf 70}, 211 (2004)
  [hep-th/0405240].

\bibitem{Grange:2005nm} 
  P.~Grange and R.~Minasian,
  ``Tachyon condensation and D-branes in generalized geometries,''
  Nucl.\ Phys.\ B {\bf 741}, 199 (2006)
  [hep-th/0512185].

\bibitem{Hatsuda:2012uk} 
  M.~Hatsuda and T.~Kimura,
  ``Canonical approach to Courant brackets for D-branes,''
  [arXiv:1203.5499 [hep-th]].

\bibitem{Koerber:2010bx} 
  P.~Koerber,
  ``Lectures on Generalized Complex Geometry for Physicists,''
  Fortsch.\ Phys.\  {\bf 59}, 169 (2011)
  [arXiv:1006.1536 [hep-th]].

\bibitem{Gualtieri:2007ng} 
  M.~Gualtieri,
  ``Generalized complex geometry,''
  [arXiv:math/0703298 [math.DG]].

\bibitem{Hitchin:2010qz} 
  N.~Hitchin,
  ``Lectures on generalized geometry,''
  [arXiv:1008.0973 [math.DG]].

\bibitem{roytenberg-1999}
  D.~Roytenberg,
  ``Courant algebroids, derived brackets and even symplectic supermanifolds,'' 
  [arXiv:math/9910078].

\bibitem{Zabzine:2006uz} 
  M.~Zabzine,
  ``Lectures on Generalized Complex Geometry and Supersymmetry,''
  Archivum Math.\  {\bf 42}, 119 (2006)
  [hep-th/0605148].

\bibitem{Bouwknegt:2010zz} 
  P.~Bouwknegt,
  ``Lectures on cohomology, T-duality, and generalized geometry,''
  Lect.\ Notes Phys.\  {\bf 807}, 261 (2010).

\bibitem{MR998124}
T.~J. Courant,
``Dirac manifolds,''
Trans.\ Amer.\ Math.\ Soc.\ {\bf 319}(2) 631--661 (1990).

\bibitem{Liu:1997fj}
Z.-J. Liu, A.~Weinstein, and P.~Xu,
``Manin triples for lie bialgebroids,''
J.\ Diff.\ Geom.\ {\bf 45} (1997) 
[dg-ga/9508013].

\bibitem{Hu}
S. Hu, ``Hamiltonian symmetries and reduction in generalized geometry,'' 
[arXiv:math/0509060 [math.DG]].

\bibitem{MR896907}
K.~Mackenzie,
{\em General theory of Lie groupoids and Lie algebroids},
London Mathematical Society Lecture Note Series 213, Cambridge University Press (2005).

\bibitem{Hull:2009mi} 
  C.~Hull and B.~Zwiebach,
  ``Double Field Theory,''
  JHEP {\bf 0909}, 099 (2009)
  [arXiv:0904.4664 [hep-th]].

\bibitem{Buscher:1987sk} 
  T.~H.~Buscher,
  ``A Symmetry of the String Background Field Equations,''
  Phys.\ Lett.\ B {\bf 194}, 59 (1987).

\bibitem{Buscher:1987qj} 
  T.~H.~Buscher,
  ``Path Integral Derivation of Quantum Duality in Nonlinear Sigma Models,''
  Phys.\ Lett.\ B {\bf 201}, 466 (1988).
 
\bibitem{Grana:2008yw} 
  M.~Grana, R.~Minasian, M.~Petrini and D.~Waldram,
  ``T-duality, Generalized Geometry and Non-Geometric Backgrounds,''
  JHEP {\bf 0904}, 075 (2009)
  [arXiv:0807.4527 [hep-th]].

\bibitem{Cavalcanti:2011wu} 
  G.~R.~Cavalcanti and M.~Gualtieri,
  ``Generalized complex geometry and T-duality,''
  arXiv:1106.1747 [math.DG].

\bibitem{Bouwknegt:2003zg} 
  P.~Bouwknegt, K.~Hannabuss and V.~Mathai,
  ``T duality for principal torus bundles,''
  JHEP {\bf 0403}, 018 (2004)
  [hep-th/0312284].

\bibitem{Myers:1999ps} 
  R.~C.~Myers,
  ``Dielectric branes,''
  JHEP {\bf 9912}, 022 (1999)
  [hep-th/9910053].

\bibitem{Giveon:1994fu} 
  A.~Giveon, M.~Porrati and E.~Rabinovici,
  ``Target space duality in string theory,''
  Phys.\ Rept.\  {\bf 244}, 77 (1994)
  [hep-th/9401139].

\bibitem{Adawi:1998ta} 
  T.~Adawi, M.~Cederwall, U.~Gran, B.~E.~W.~Nilsson and B.~Razaznejad,
  ``Goldstone tensor modes,''  JHEP {\bf 9902}, 001 (1999)  [hep-th/9811145].

\bibitem{Rey:1989ti} 
  S.~-J.~Rey,
  ``The Higgs Mechanism For Kalb-ramond Gauge Field,'' 
  Phys.\ Rev.\ D {\bf 40}, 3396 (1989).

\bibitem{Yokoi:2000en} 
  N.~Yokoi,
  ``Nonlinear realization of Lorentz symmetry,''
  Phys.\ Lett.\ B {\bf 504}, 109 (2001)
  [hep-th/0011158].

\bibitem{Higashijima:2001sq} 
  K.~Higashijima and N.~Yokoi,
  ``Spontaneous Lorentz symmetry breaking by antisymmetric tensor field,''
  Phys.\ Rev.\ D {\bf 64}, 025004 (2001)
  [hep-th/0101222].

\bibitem{Koerber:2005qi} 
  P.~Koerber,
  ``Stable D-branes, calibrations and generalized Calabi-Yau geometry,''
    JHEP {\bf 0508}, 099 (2005)
  [hep-th/0506154].

\bibitem{Martucci:2005ht}
  L.~Martucci and P.~Smyth,
  ``Supersymmetric D-branes and calibrations on general N=1 backgrounds,''  JHEP {\bf 0511}, 048 (2005)  [hep-th/0507099].  

\bibitem{Martucci:2006ij} 
  L.~Martucci,
  ``D-branes on general N=1 backgrounds: Superpotentials and D-terms,''  JHEP {\bf 0606}, 033 (2006)  [hep-th/0602129].  

\bibitem{Koerber:2006hh} 
  P.~Koerber and L.~Martucci,
  ``Deformations of calibrated D-branes in flux generalized complex manifolds,''
  JHEP {\bf 0612}, 062 (2006)
  [hep-th/0610044].

\bibitem{Lust:2010by} 
  D.~Lust, P.~Patalong and D.~Tsimpis,
  ``Generalized geometry, calibrations and supersymmetry in diverse dimensions,''
  JHEP {\bf 1101}, 063 (2011)
  [arXiv:1010.5789 [hep-th]].

\bibitem{Jeon:2011cn} 
  I.~Jeon, K.~Lee and J.~-H.~Park,
  ``Stringy differential geometry, beyond Riemann,''
  Phys.\ Rev.\ D {\bf 84}, 044022 (2011)
  [arXiv:1105.6294 [hep-th]].

\bibitem{Jurco:2012yv} 
  B.~Jurco and P.~Schupp,
  ``Nambu-Sigma model and effective membrane actions,''
  [arXiv:1203.2910 [hep-th]].

\bibitem{Schupp:2012nq} 
  P.~Schupp and B.~Jurco,
  ``Nambu Sigma Model and Branes,''
  [arXiv:1205.2595 [hep-th]].

\bibitem{Gualtieri:2007bq} 
  M.~Gualtieri,
  ``Branes on Poisson varieties,''
  [arXiv:0710.2719 [math.DG]].

\bibitem{Asakawa:2001vm}
  T.~Asakawa, S.~Sugimoto and S.~Terashima,
  ``D-branes, matrix theory and K homology,''  
  JHEP {\bf 0203} (2002) 034
  [hep-th/0108085].  

\end{thebibliography}
\end{document}